\documentclass{aa}
\usepackage{natbib}\usepackage{txfonts}
\usepackage{graphicx}

\newcommand{\lp}{ \;\left(}
\newcommand{\rp}{ \right)}
\newcommand{\nab}{ \vec{\nabla} }
\newcommand{\lapl}{ \Delta }

\newcommand{\noi}{ \noindent }

\newcommand{\beq}{\begin{equation}}
\newcommand{\eeq}{\end{equation}}

\newcommand{\lc}{ \left[}
\newcommand{\rc}{ \right]}
\newcommand{\greq}{\begin{equation} \begin{array}{l}}
\newcommand{\egreq}{\end{array} \end{equation}}

\newcommand{\YL}{ Y^m_\ell }
\newcommand{\YLM}{ Y^{m'}_\ell } 
\newcommand{\YTL}{\hat{Y}^m_\ell}

\newcommand{\eeqn}[1]{\label{#1}\end{equation}}

\newcommand{\eq}[1]{(\ref{#1})}

\newcommand{\beqa}{\begin{eqnarray*}}
\newcommand{\eeqa}{\end{eqnarray*}}
\newcommand{\beqan}{\begin{eqnarray}}
\newcommand{\eeqan}[1]{\label{#1}\end{eqnarray}}

\begin{document}

\title{Asymptotic analysis of high-frequency acoustic modes in rapidly rotating stars}

\author{F. Ligni\`eres \inst{1} \and B. Georgeot \inst{2,3}}
\institute{
Laboratoire d'Astrophysique de Toulouse-Tarbes,
Universit\'e de Toulouse, CNRS,
31400 Toulouse,
France \\
\email{francois.lignieres@ast.obs-mip.fr}
\and
Universit\'e de Toulouse; UPS; Laboratoire de
Physique Th\'eorique (IRSAMC); F-31062 Toulouse, France
\and
CNRS; LPT (IRSAMC); F-31062 Toulouse, France}

\offprints{F. Ligni\`eres}

\date{Received  \today / Accepted ??}

\abstract
{The asteroseismology of rapidly rotating pulsating stars is hindered by our poor knowledge of the effect of the rotation
on the oscillation properties.}
{Here we present an asymptotic analysis of high-frequency acoustic modes in rapidly rotating stars.}
{We study the Hamiltonian dynamics of acoustic rays in uniformly rotating polytropic stars and show that the phase
space structure has a mixed character,
regions of chaotic trajectories coexisting with stable structures like island chains or invariant tori.
In order to interpret the ray dynamics in terms of acoustic mode properties, we then use tools and concepts
developed in the context of quantum physics.} 
{Accordingly, the high-frequency acoustic spectrum is a
superposition of frequency subsets associated with
dynamically independent phase space regions. The sub-spectra associated with stable structures are regular
and can be modelled through EBK quantization methods while those associated with chaotic regions are irregular but with generic
statistical properties.
The results of this asymptotic analysis are successfully confronted with the properties of numerically computed 
high-frequency acoustic modes. The implications for the asteroseismology of rapidly rotating stars are discussed.}
{}


\maketitle

\section{Introduction}

Interpreting the oscillation spectra of rapidly rotating stars is a long standing unsolved problem of asteroseismology.
It so far prevented to directly probe the interior of stars in large fractions of the HR diagram, mostly in the range of intermediate and massive stars.
Important progresses are nevertheless expected from the current spatial seismology missions (in particular COROT and KEPLER)
as well as from recent modelling efforts on the effect of rotation on stellar oscillations.
On the modelling side, new numerical codes have been able to accurately compute oscillation frequencies
in significantly centrifugally distorted polytropic models of stars \citep{Lign,Rees} and are being extended to more realistic models \citep{Rees09}.
In particular the previous calculations based on perturbative methods were shown
to be inadequate for these stars \citep{Rees,Lovekin}.
Nevertheless, interpreting the data requires an understanding of the mode properties that goes far beyond the
accurate computation of frequencies. Indeed, the necessary identification of the observed frequencies with
theoretical frequencies is a largely underconstrained problem which requires a priori information on the spectrum to be successful.
The knowledge of the structure of the frequency spectrum is of primary importance in this context.
For slowly rotating pulsating stars, this structure is characterized by regular frequency patterns which can be analytically derived
from an asymptotic theory of the high-frequency acoustic modes.

Until recently, the asymptotic structure of the frequency spectrum of rapidly rotating stars
was completely unknown.
Our first calculations of low degree ($\ell =0-7$) and low order ($n=1-10$) acoustic axisymmetric modes in centrifugally distorted polytropic stars
\citep{Lign}
have revealed the presence of regular frequency patterns similar albeit different from those
of spherically symmetric stars.
This was confirmed using more realistic calculations including the Coriolis force and was also extended
to non-axisymmetric and higher-frequency modes \citep{Rees8}.
The analogy with the non-rotating case suggests an asymptotic analysis could model these empirical regular patterns.

The asymptotic analysis presented in this paper is based on the acoustic
ray dynamics. This approach can be viewed as a natural extension of the asymptotic analysis developed for non-rotating stars \citep{Vandakurov,Tas80, 
Tas90, Deub, Roxb}.
In this case, spherical symmetry enables to reduces the initial 3D boundary value problem to a 1D boundary value problem
in the radial direction.
Asymptotic solutions of this 1D boundary value problem can then be obtained
using a short-wavelength approximation
which consists in looking for wave-like solutions
under the assumption that
their wavelength is much smaller
than the typical lengthscale of the background medium.
As rotation breaks the spherical symmetry, the eigenmodes are no longer separable in the latitudinal and radial directions and the 3D
boundary value problem of acoustic modes in rapidly rotating stars can not be reduced to a 1D boundary value problem.
Thus, the short-wavelength approximation is directly applied to the 3D equations governing linear adiabatic stellar perturbations.
This leads to an acoustic ray model which describes the propagation of locally plane waves.
Since we are concerned by oscillation modes,
the main issue of an asymptotic analysis based on ray dynamics
is to construct standing waves solutions from the
short-wavelength propagating waves described by the acoustic rays.

The short-wavelength approximation of wave equations is standard in physics, best
known examples being the geometrical optics limit of
electromagnetic waves or the
classical limit of quantum mechanics.
We shall see that similarly to these cases
the acoustic rays in stars can be described
as trajectories of a particle in the framework of classical Hamiltonian mechanics.
As is well known in quantum physics \citep{Gutzwiller, Ott}, the issue of construction modes from the ray dynamics
depends crucially on the nature of this Hamiltonian motion.


Indeed, Hamiltonian systems can have one of two very
different behaviors.  If there are enough constants of motion, the
dynamics is integrable, and trajectories organize themselves in
families on well-defined phase space structures.  In contrast, chaotic
systems display exponential divergence of nearby trajectories, and a
typical orbit is ergodic in phase space. The modes constructed from these
different dynamics are markedly different.
For integrable systems, the eigenfrequency spectrum
can
be described by a function of $N$ integers, $N$ being the number of degrees of freedom of the system.
In contrast, the spectrum of chaotic systems shows no such regularities. However, the spectrum possesses
generic statistical properties which can be predicted.
In the last thirty years, the field called quantum chaos
has developed concepts and methods to relate non-integrable ray
dynamics to properties of the associated quantum systems (and more generally wave systems).

We shall see that the acoustic ray dynamics in rotating stars undergoes a transition from an integrable system
at zero rotation to a mixed system, that is a system with a phase space containing integrable
and chaotic regions, at high rotation.
The acoustic ray dynamics of rotating stars being non-integrable, we are led to
use quantum chaos theory to predict the asymptotic properties of acoustic modes.

In the following, we shall describe in detail the ray dynamics, the predictions on the modes properties
and their validation through a comparison with numerically computed acoustic modes.
But, before going into the technical details of this analysis, we would like to summarize our results,
giving emphasis to those of practical interest for mode identification.
These results have been obtained for a sequence of uniformly rotating polytropic models, but we expect them to
be qualitatively correct for a wider range of models.
At high rotation rates, the frequency spectrum can be generically described as the superposition of
an irregular frequency subset and of various regular frequency subsets each showing specific
patterns.
This spectrum structure is significantly more complex that in the spherical case where all acoustic
frequencies follow the same organization such as given by Tassoul's formula \citep{Tas80}.
However, in the observed spectrum, only two frequency subsets are expected to be dominant.
One subset (the subset of island modes) shows regular patterns similar albeit different to those found in non-rotating stars (it corresponds to
the modes subset studied by \citet{Lign,Rees8}).
These modes are associated with rays whose dynamics is near-integrable and consequently asymptotic formulas describing their regular patterns can be derived.
The second frequency subset (the subset of chaotic modes) is associated with chaotic rays. Although it does not follow a regular pattern, specific statistical properties
of this frequency subset can be predicted.
Besides, the asymptotic analysis provides an estimate of the proportion of mode in each subset.
The transition from the non-rotating case occurs as follows: At
moderate rotation, the regular subset of island modes is superposed to another
regular subset (the subset of whispering gallery modes)
which is a direct continuation of the non-rotating spectrum. At this stage, chaotic modes are rare and difficult to observe. As rotation increases, the
number of chaotic modes increases while whispering gallery modes become less and less visible.
Obviously, such a priori informations on the frequency spectrum should be of practical interest for the mode identification.
One should however keep in mind that the asymptotic analysis is not supposed to be accurate for the
lowest frequency p-modes. Although a detailed study of the asymptotic analysis limit of validity has not been performed yet,
numerical results indicate that the regular patterns are quite accurate down to $5^{\rm th}$ radial order \citep[see][]{Lign,Reest}.
At lower radial orders, the asymptotic mode classification in different subsets could still be applicable.

The paper is organized as follows: The equations governing the star model, the small perturbations about this model,
the short-wavelength approximation of
these perturbations and the ray model for progressive acoustic waves are presented in Section 2.
A detailed numerical study of the acoustic ray dynamics has been conducted for uniformly rotating polytropic models of stars.

The results are analyzed in Section 3 using Poincar\'e Surface of
Section, a standard tool
to visualize the phase space structure. It shows that, as rotation increases, the dynamics undergoes a transition from integrability (at zero rotation) to a mixed state
where parts of the phase space display integrable
behaviour and while other parts are chaotic.

We then relate the acoustic ray dynamics to the asymptotic properties of the acoustic modes (Section 4).
We first summarize the results obtained in the context of quantum physics distinguishing the cases where the
Hamiltonian system is integrable, fully chaotic or of mixed nature. In accordance with this theory,
we show that the asymptotic acoustic spectrum of the uniformly rotating polytropic models of stars
is a superposition of regular frequency patterns and
irregular frequency subsets respectively associated with near-integrable and
chaotic phase space regions. The averaged number of modes associated with each phase space regions
depends directly
on its
volume (in phase space).
These predictions are then successfully confronted with the actual properties
of high-frequency acoustic modes computed for a particular fast
rotating stellar model

In Section 5, after a critical discussion of the assumptions of the
asymptotic analysis,
we show how our results can be used for the mode identification
and  for the seismic studies of rapidly rotating stars. The conclusion is given in section 6.

The present work complements and extends a short recent paper \citep{pre} by giving all the necessary
details of the analysis and by presenting new results concerning (i) the ray dynamics at different rotation
rates and
for non-vanishing values of the angular momentum projection onto the
rotation axis $L_z$,
(ii) the analysis of extra regular patterns visible for some specific
values of rotation,
 (iii) the number of modes in each frequency subset
predicted by the asymptotic analysis, (iv) the visibility of the chaotic modes.

\section{Formalism and numerical methods} 

In this section we present the equations governing the star model (subsection 2.1), the small perturbations about this model (subsection 2.2), the short-wavelength approximation of
these perturbations (subsection 2.3) and the ray model for progressive acoustic waves (subsection 2.4).
The numerical method used to compute the ray trajectories is described in subsection 2.5.
The oscillation modes have been computed with the code 
described in \citet{Lign} and \citet{Rees}.

\subsection{Polytropic model of rotating star}

The model is
a self-gravitating uniformly rotating monatomic gas ($\Gamma = 5/3$) verifying a polytropic relation
assumed to give a reasonably
good approximation
of the relation between the pressure and the density in the star \citep{Hansen}:
\begin{eqnarray} \label{hyd}
P_0 &=& K \rho_0^{1+1/\mu} \\
0 &=& -\nab P_0 - \rho_0 \nab \left( \psi_0 -\Omega^2 w^2/2 \right)  \\
\lapl \psi_0 &=& 4\pi G\rho_0
\end{eqnarray}
where $P_0$ is the pressure, $\rho_0$ the density, $K$ the polytropic constant,
$\mu$ the polytropic index, $\psi_0$ the gravitational potential, $\Omega$ the rotation rate,
$w$ the distance to the
rotation axis and $G$ the gravitational constant.

The uniform rotation ensures that the
fluid is barotropic.
A pseudo-enthalpy can then be introduced $h_0=\int dP_0/\rho_0 = (1+\mu)P_0/\rho_0$
and the integration of the hydrostatic equation reads:
\begin{equation}
\label{eq:hydrostatic.enthalpy}
h_0 = h_c -(\psi_0 - \psi_c) + \frac{1}{2}\Omega^2 w^2
\end{equation}
where the subscript ``$c$'' denotes the value in the center of the polytropic model.
Equation~\eq{eq:hydrostatic.enthalpy} is then inserted into Poisson's equation to yield:
\begin{equation}
\label{eq:Poisson.phi}
\lapl \psi_o = 4 \pi G \rho_c \left( 1 - \frac{\psi_o - \psi_c}{h_c} +
\frac{\Omega^2 d^2}{2 h_c} \right)^{\mu}
\end{equation}

Equation~\eq{eq:Poisson.phi} is solved numerically, using an iterative scheme,
as described in \citet{Ri05}. 

Specifying the mass and the rotation rate of the star is not sufficient to determine
the polytropic model in physical units. 
This requires to fix an additional parameter, for example the stellar radius (see \citet{Hansen} for the non-rotating case and 
\citet{Chris0} for a brief discussion 
on a suitable parameter choice for rotating stars). In the following, however, 
we only present dimensionless quantities which do
not depend on the choice of this additional parameter.
The rotation rate $\Omega$ is compared to $\Omega_K = \left(GM/R_e^3\right)^{1/2}$
the limiting rotation rate for which the centrifugal acceleration equals the gravity at the equator,  $M$ being the stellar mass and $R_e$ the equatorial
radius. The star flatness is  $\epsilon= 1 -R_p /R_e$ where $R_p$ is the polar radius. The acoustic frequencies shall be expressed in terms
of $\omega_0 = \left(GM/R_p^3\right)^{1/2}$, the inverse of a dynamical time scale, 
or $\omega_1$ the lowest acoustic mode frequency of the stellar model considered.

\subsection{Perturbation equations and boundary conditions}

Time-harmonic small amplitude perturbations of the star model are studied under two basic 
assumptions. The first is to neglect the Coriolis force. This a natural assumption
to study high-frequency acoustic modes since the oscillation time scale
is asymptotically much shorter than the Coriolis force time scale $1/(2\Omega)$.
Moreover, complete calculations by \citet{Rees} (see figure 6 of this paper) have shown that
Coriolis force effects on the frequency are already very small for relatively low 
radial order (say $n \approx 5$).
The second basic assumption is to neglect the viscosity and the non-adiabatic effects.
This is a standard approximation in the asymptotic analysis since these effects have little influence on the
value of the frequency.
Both assumptions shall have important consequences on the acoustic ray dynamics described below.
Neglecting the Coriolis force ensures that the dynamics is symmetric with respect to
the time reversal while the absence of diffusion processes makes the dynamics conservative.
Finally, the Cowling approximation valid for high frequencies enables
to neglect the perturbation of the gravitational potential.
Under these assumptions, the linear equations governing the evolution
of small amplitude perturbations read:

\begin{equation} \label{div}
{\partial}_t \rho + \nab \cdot (\rho_{0} \vec{u})=0,
\end{equation}

\begin{equation} \label{vel}
\rho_{0} {\partial}_t \vec{u} =
- \nab P + \rho \vec{g}_0, 
\end{equation}

\begin{equation} \label{adia}
{\partial}_t P + \vec{u} \cdot \nab P_0 = c_s^{2} \lp {\partial}_t \rho + \vec{u} \cdot \nab \rho_0 \rp,
\end{equation}

\noi where $\vec{u}$, $\rho$ and $P$, are respectively the Eulerian perturbations of velocity,
density and pressure. The sound speed is
$c_s = \sqrt{{\Gamma} P_0 / \rho_{0}}$, ${\Gamma}$ being the first adiabatic exponent of the gas,
and $\vec{g}_0 = - \nab \left( \psi_0 -\Omega^2 w^2/2 \right)$ is the effective gravity.

As in \citet{Pe38}, because the pressure and the temperature of the stellar model is zero at
the surface, 
the only condition to impose on the perturbations is to be regular everywhere.

\subsection{The short-wavelength approximation of the perturbation equations}

The acoustic ray model results 
from a short-wavelength approximation of the perturbation equations \eq{div}, \eq{vel}, \eq{adia}, called the Wentzel-Kramers-Brillouin (WKB) approximation.
Time-harmonic wave-like solutions
of the form,
\begin{equation} \label{eq:WKB}
\Psi = \Re\lbrace A({\bf x}) \exp [i \Phi ({\bf x}) - i \omega t] \rbrace = \Re\lbrace \hat{\Psi}({\bf x}) \exp (- i \omega t)\rbrace
\end{equation}
are sought under the assumption that
their wavelength is much smaller
than the typical lengthscale of the background medium.
As discussed by \citet{Gough}, one expects a better approximation if the starting equations \eq{div}, \eq{vel}, \eq{adia}
are first reduced to a so-called normal form which avoids first order derivatives. This is done in Appendix A.1 leading to:
\begin{equation} \label{pert}
\frac{\omega_c^2 - \omega^2}{c_s^2} \hat{\Psi} + \frac{N_0^2}{\omega^2} [\Delta - \frac{1}{g_0^2}(\vec{g_0} \cdot \nab) (\vec{g_0} \cdot \nab)] \hat{\Psi} = \Delta \hat{\Psi}
\end{equation}
\noi where $\hat{\Psi} = \hat{P}/\alpha$, $\hat{P}$ is the complex amplitude associated with the pressure perturbation
$P=\Re\lbrace\hat{P} \exp (- i \omega t)\rbrace$
and $\alpha$ is a function of the background star model given by Eq.~\eq{alpha}.
The expressions of the cut-off frequency $\omega_c$ and the Brunt-V\"{a}is\"{a}l\"{a} frequency $N_0$
are given respectively by \eq{omcgene} and \eq{Brunt}.
Note that the two left-hand-side terms of equation \eq{pert} account respectively for acoustic and gravity waves.

As detailed in  Appendix A.2, the WKB approximation is then applied to \eq{pert}. The dominant term of the expansion in powers
of the ratio between the wavelength solution and the background typical lengthscale yields
an equation governing the phase $\Phi ({\bf x})$, the so-called eikonal equation. 
The amplitude $A({\bf x})$ is determined at the next order as a function of $\Phi ({\bf x})$.
Neglecting the gravity waves by taking the high frequency limit, the eikonal equation reads:
\begin{equation} \label{wkb2}
\omega^2 = \omega_c^2 + c_s^2 k^2
\end{equation}
\noi where $\vec{k}= \nab \Phi$ is the wavevector.
Moreover, for a polytropic model of star and using the approximation $\omega \gg  N_0$ valid for high 
frequency acoustic modes,
the function $\alpha$ is proportional to $\rho_{0}^{1/2}$ and the cut-off frequency $\omega_c$ is simplified to:
\begin{equation} \label{om1dd}
\omega_c^2 = \lc \frac{\Gamma \mu (\mu+2)}{2 (\mu+1)} -2 \rc \frac{g_0^2}{2 h_0} + \frac{1}{2} \lp 2 - \frac{\Gamma \mu}{\mu+1}\rp \nab \cdot \vec{g}_0,
\end{equation}
In the range of high frequency acoustic modes, 
the $\omega_c$ term is expected to be much smaller than $\omega$ in
most parts of the star except near the surface where $\omega_c$ diverges.
Note also that, despite the chosen notation, $\omega_c^2$ can be negative near the center.

\subsection{The acoustic ray model as a Hamiltonian system}

The acoustic ray model consists in finding solutions of the eikonal equation \eq{wkb2} along some path called the ray path.
This problem can be written in a Hamiltonian form
where the solutions $(\vec{x}(t), \vec{k}(t))$ are conjugate variables of the Hamiltonian and $t$, the parameter
along the path, is a time-like variable.
To derive the Hamiltonian equations from the eikonal equation, one can apply a procedure valid for a general 
eikonal equation $D(\vec{k}, \omega, \vec{x}) =0$ \citep[e.g.][]{Ott}. This leads to the Hamiltonian $H'=\omega=\sqrt{c_s^2 {\bf k}^2 + {\omega}_c^2}$ \citep[e.g.][]{Light}.
Another useful Hamiltonian formulation can be readily obtained by considering a path normal to the wavefront $\Phi(\vec{x}) = const.$
(this method being also used to determine optical rays in isotropic media of variable index \citep{Born}, the quantity $(\sqrt{1-\omega_c^2/\omega^2})/c_s$
playing
the role of the medium index $n(\vec{x})$).
The ray path is thus defined by
\begin{equation} \label{path}
\frac{d \vec{x}}{d s} = \frac{\vec{k}}{\| \vec{k} \|} = \frac{\nab \Phi}{\| \nab \Phi \|}
\end{equation}
\noi where $s$ is the curvilinear coordinate
along the ray. Differentiating \eq{path} and using \eq{wkb2} then leads to the following system of ODEs:
\begin{eqnarray} \label{prop}
\frac{d \vec{x}}{d t} &=& \vec{\tilde{k}} \\
\label{prop2}
\frac{d \vec{\tilde{k}}}{d t} &=& - \nab W
\end{eqnarray}
\begin{equation} \label{pot}
W = - \frac{1}{2 c_s^2} \lp 1 - \frac{\omega_c^2}{\omega^2} \rp
\end{equation}
\noi where we use the frequency-scaled wavevector
$\vec{\tilde{k}} = \vec{k} / \omega$ and the time-like variable $t$ such that $d t = c_s d s / (1-  \omega_c^2 / \omega^2)^{1/2}$.

As $W$ only depends on the spatial variable $\vec{x}$, the second equation has the classical form of Newton's second law
for the conservative force associated with the potential $W$
(for a unit mass and a time variable $t$). This system can thus be written in a Hamiltonian form where
\begin{equation} \label{ham}
H = \frac{\vec{\tilde{k}}^2}{2} + W(\vec{x})
\end{equation}
\noi is the Hamiltonian. According to the eikonal equation \eq{wkb2}, $H$ is equal to zero
and the dynamics is therefore fully determined by the potential well $W$, where the frequency $\omega$ appears as a parameter. 
As $\omega \gg {\omega}_c$ away from the near surface layers, the potential increase towards the star center is given by the sound speed
increase. Close to the surface, the potential barrier is due to the sharp increase of $\omega_c$ and provokes
the wave back-reflection. While
the location of the reflection surface depends on the frequency $\omega$, we shall see that the dynamics 
is not strongly dependent on $\omega$ in the range of high-frequency acoustic modes
here considered. This can be expected since, as $\omega_c$ diverges towards the surface of the polytropic model of star, 
the location of $\omega=\omega_c$ does not vary rapidly with $\omega$.

The potential being symmetric with respect to the rotation axis of the star, the angular momentum projection on this axis
$\tilde{L}_z = r \sin \theta \tilde{k}_{\phi}$ is a constant of motion, where $\tilde{k}_{\phi} = \vec{\tilde{k}} \cdot \vec{e}_{\phi}$
and $\vec{e}_{\phi}$ is a unit vector in the azimuthal direction and $[r,\theta,\phi]$ are the spherical coordinates. 
Consequently, the projection of the ray trajectory on the meridional plane rotating with the ray at an angular velocity
$d \phi / dt = \tilde{L}_z / (r \sin \theta)^2$ is governed by the two-degree-of-freedom Hamiltonian:
\begin{equation} \label{ham2d}
H_r = \frac{\vec{\tilde{k}_p}^2}{2} + W_r(\vec{x})
\end{equation}
where $\vec{\tilde{k}_p} = \vec{\tilde{k}} - \tilde{k}_{\phi} \vec{e}_{\phi}$ is the wavevector projected onto the rotating meridional plane
and $W_r$ is the effective potential of the reduced Hamiltonian $H_r$ which
now also depends on $\tilde{L}_z$ as a parameter:
\begin{equation} \label{pot2d}
W_r(\vec{x})=\frac{\tilde{L}_z^2}{2(r \sin \theta)^2} - \frac{1}{2 c_s^2} \lp 1 - \frac{\omega_c^2}{\omega^2} \rp
\end{equation}

\subsection{Numerical method for the ray dynamics}

The acoustic ray dynamics has been investigated by integrating numerically Eqs.~\eq{prop} and \eq{prop2} using
a 5th order Runge-Kutta method. The step size of the integration is chosen automatically
to keep the local error estimate smaller than a given tolerance.
To what extent this control of the local error ensures that the numerical solution is
close to the true
 solution depends on the stability of the problem.
For chaotic trajectories, the numerical error tends to grow rapidly while
for stable trajectories this error remains of the same order as the relative error.  The rapid growth of numerical errors
is one of the characteristics of chaotic dynamics; however, this does not
prevent to simulate such systems since for hyperbolic systems
the shadowing theorem \citep{shadowing,sauer} ensures
that an exact trajectory will remain close to the dynamics of each
computed point for arbitrary times. Thus while a numerical trajectory
diverges from the exact one,
it nevertheless remains close to another exact trajectory, and therefore
numerical errors do not prevent to obtain accurate phase space plots.
We checked that the Poincar\'e surface of section shown in the next section 
are not significantly modified by decreasing the maximum allowed local error.
We also checked the influence of the resolution of the background polytropic stellar model.
Increasing this resolution from 62 to 92 Gauss-Lobatto points in the pseudo-radial direction has no
significantly  effect on the Poincar\'e surface of section.
Finally, the Hamiltonian conservation is used as an independent accuracy test
on the
computation.

\section{Acoustic ray dynamics in rotating stars}

In this section, we show that rotation strongly modifies the acoustic ray dynamics.
Indeed, we find that, as rotation increases, the dynamics undergoes a transition from integrability (at zero rotation) to
a mixed state
where parts of the phase space display integrable
behaviour while other parts are chaotic.

The nature of a dynamical system is best investigated by considering the structure of its phase space which contains
both position and momentum coordinates. We thus
first introduce the Poincar\'e Surface of Section (hereafter the PSS) which is a standard tool to visualize the phase
space (subsection 3.1).
Then we describe the phase space structure
at zero rotation (subsection 3.2) and the main features of the generic phase space structure
at high rotation rates (subsection 3.3). The detail of the transition to chaos as rotation increases is analyzed in subsection 3.4. As this last
subsection makes use of several specific tools and theorems of dynamical systems theory, it might be skipped at first reading.

\subsection{Phase space visualization : The Poincar\'e surface of section}

As shown in subsection 2.4, acoustic rays with a given $L_z$ are governed by a Hamiltonian with two degrees of freedom $H_r$. 
The associated phase space is
therefore four-dimensional and difficult to visualize.  
A PSS is constructed by computing the intersection of the phase space trajectories
with a chosen ($2N-1$)-dimensional surface, where $N$ is the number of degrees of freedom of the system.
If $H$ is time-independent, then energy conservation implies that phase space trajectories stay on a ($2N-1$)-dimensional surface.
The 
PSS is thus a ($2N-2$)-dimensional surface in general and a $2$-dimensional surface in the present case.

\begin{figure}[htb]
\resizebox{\hsize}{!}{\includegraphics{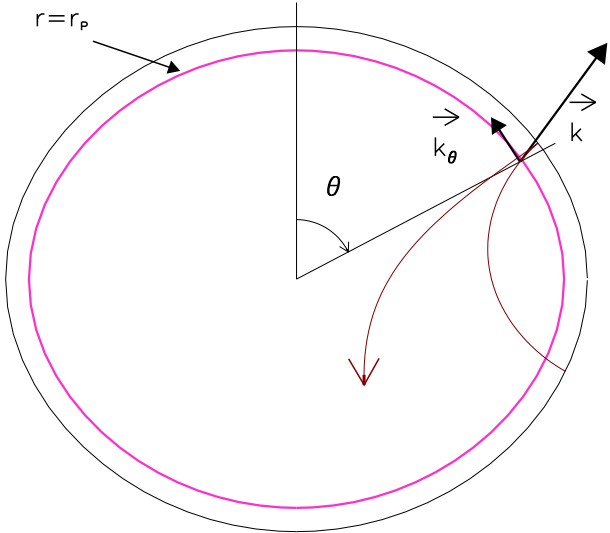}}
\caption{(Color online) Intersection of an outgoing acoustic ray (red/arrow headed)  with the $r = r_p(\theta)$ curve (magenta/thick).
The point on the associated PSS is specified
by the colatitude $\theta$ and the scaled latitudinal wavenumber component $k_{\theta}/\omega$ at the intersection.}
\label{fig0}
\end{figure}

Different choices are possible for the PSS although some conditions are required to obtain a 
good description of the dynamics \citep[see for example][]{Ott}.
First, in order to provide a complete view of phase space, the PSS must be intersected by all phase space trajectories. Here we chose
the curve $r_p (\theta)=r_s (\theta) - d$
situated at a fixed radial distance $d$ from the stellar surface $r_s (\theta)$ displayed on Fig.~\ref{fig0}.
As shown in Appendix B.1
for the non-rotating case,
the distance $d$ can be chosen such that all relevant trajectories intersect this curve.
The second condition is that, given a point on the PSS, the next point on the PSS has to be 
uniquely determined. This is ensured by counting the intersection with $r_p (\theta)=r_s (\theta) - d$
only when the
trajectory comes from on side of the
$r=r_p(\theta)$ curve (here we consider the trajectories coming from the inner side).
Finally, the coordinate system used to display the PSS is chosen in order that any surface of the PSS is conserved by the dynamics in the same way as four-dimensional volumes are preserved in
phase space. The coordinates $[ \theta, \tilde{k}_{\theta} ]$ where
$\tilde{k}_{\theta}$ is the latitudinal component of $\vec{\tilde{k}}$
in the
natural basis $(\vec{E}^{\zeta},\vec{E}^{\theta},\vec{E}^{\phi})$ associated with the coordinate system $[\zeta=r_s (\theta)-r, \theta, \phi]$
fulfill this condition (as shown in Appendix B.2).

\begin{figure*}
\resizebox{\hsize}{!}{\includegraphics{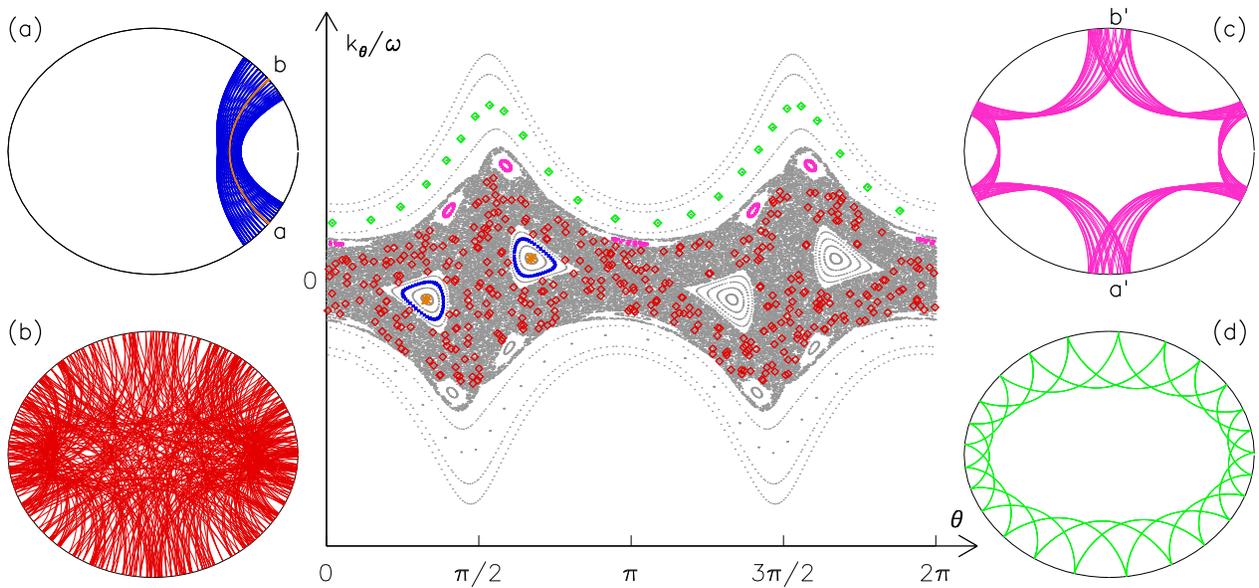}}
\caption{(Color online) PSS at $\Omega = 0.59 \Omega_K$ and typical acoustic rays associated with
the four main
phase space structures : (a) a 2-period island ray (blue/dark grey) and the associated periodic orbit
with endpoints $a$ and $b$
(orange/light grey), (b) a chaotic ray (red/grey), (c) a 6-period island ray (magenta/light grey)
and (d) a whispering gallery ray (green/light grey).
On the PSS, (colored/grey) symbols (diamonds for the chaotic and whispering gallery rays, crosses for the 2-period and 6-period island rays) specify the po
ints where these trajectories cross the PSS.}
\label{fig3}
\end{figure*}

PSS have been obtained by following many trajectories of different initial conditions.
The number of trajectories together with the time during which they are computed
determine the resolution by which the phase space is investigated.
In principle, we should display PSS computed for different values of the frequency $\omega$.
However, as $\omega$
is varied in the range of frequency here considered, we found that the PSS remained practically unchanged.
As discussed in subsection 2.4, this stems from the fact 
that the dynamics of the frequency-scaled wavevector $\vec{k}/\omega$ is weakly dependent on $\omega$ in this frequency range.


\subsection{The non-rotating case $\Omega=0$}

\begin{figure}[htb]
\resizebox{\hsize}{!}{\includegraphics{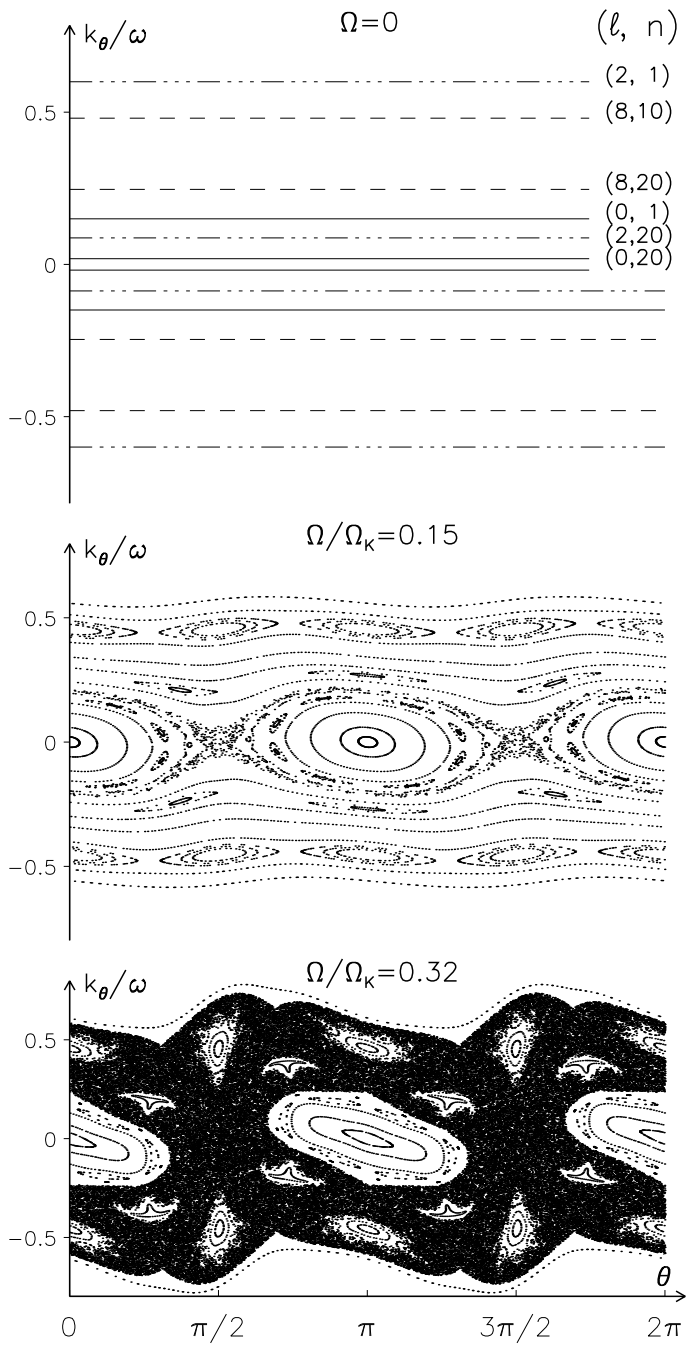}}
\caption{Three $\tilde{L}_z=0$ PSS at small rotation rates showing the transition to chaos. The unit of $k_{\theta}/\omega$ is $\omega_0^{-1}$.
Island chains and chaotic regions
appear around respectively stable and unstable periodic orbits. On the $\Omega= 0$ PSS, the straight lines correspond to
intersections with mode-carrying-tori specified
by the degree and radial order of the mode.}
\label{fig1}
\end{figure}

The PSS at $\Omega=0$ is described in this subsection. It will serve as a reference to
investigate the evolution of the dynamics with rotation.
Due to spherical symmetry, the norm of the angular momentum with respect to the star center
\begin{equation} \label{degree}
\tilde{L}=\sqrt{\tilde{k}_{\theta}^2
+ \left( \frac{\tilde{L}_z}{\sin \theta} \right)^2 }
\end{equation}
\noi is a conserved quantity. It follows that the intersection of any trajectory with the PSS belongs to
a curve of the form:
\begin{equation} \label{pssint}
\tilde{k}^2_{\theta}= \tilde{L}^2 -\left( \frac{\tilde{L}_z}{\sin \theta} \right)^2
\end{equation}
\noi For $\tilde{L}_z=0$, these are the two straight lines $\tilde{k}_{\theta}=\pm \tilde{L}$
(see Fig.~\ref{fig1}) while Eq.~\eq{pssint} yields a closed curve for $\tilde{L}_z \neq 0$ the trajectories being constrained to
latitudes smaller than $\arcsin(|\tilde{L}_z |/\tilde{L})$ (see Fig.~\ref{fig2.lz5}). This curve varies from a near rectangle
to an ellipse as $\tilde{L}_z$ grows from $0^+$ to $\tilde{L}$.

The simplicity of the PSS reflects the fact that the system is integrable (\eq{degree} indeed provides the second 
invariant (in addition to $H_r$) 
of the reduced two-degree-of-freedom dynamical system). 
Every integrable system is structured in $N$-dimensional surfaces which are associated with specific values
of the $N$ constants of motion. This means that any trajectory is constrained to stay forever on one of these surfaces.
They are called invariant tori because they are invariant through the dynamics and they have a torus-like topology.
As we shall verify in the following,
they play a crucial role in 
the transition to chaos as well as in the modes construction.
The PSS at $\Omega=0$ actually visualize the intersection of these tori with the $r = r_p$ surface.

Importantly, the invariant tori
can be of two different types which determine their fate once the rotation is increased.
Rational (or resonant) tori are such that
all trajectories on the torus are periodic orbits (that is trajectories that close on themselves in phase space).
In contrast, irrational tori are such that any trajectory densely covers the whole torus.


\subsection{Phase space structure at high rotation rates}

The main features of the phase space at high rotation rates are shown in Fig.~\ref{fig3}
where the PSS at $\Omega = 0.59 \Omega_K$ is displayed together with four acoustic rays shown
on the position space as well as on the PSS.
These rays belong to the three different types of phase space structures always present at high rotation rates.
First, a large chaotic region appears (the grey region on Fig.~\ref{fig3}). Chaotic rays, e.g. the red ray, are not constrained to stay on tori (that is on one-dimensional curves on the PSS) and thus fill up
densely and ergodically a phase space volume (that is a surface on the PSS). 
Second, the island chain embedded in the large chaotic region is a common structure of phase space at high rotation rate.
An important property of the island chain is to be structured by invariant tori centered on the
periodic orbit of period 2 (the orange ray).
The PSS also shows smaller island chains
like the one formed around a 6-period periodic orbit (see the magenta ray). However, contrary to the 2-period island chain, such 
structures survive only up to a certain rotation rate.
Third, the undulated curves present in the high $\tilde{k}_{\theta}$ region are formed by whispering gallery type trajectories (like
the green ray),
that is trajectories following the outer boundary. The associated tori correspond to the deformation of non-rotating tori which have not been destroyed 
yet at this rotation rate.
Overall this type of phase space organization is typical of mixed systems
with coexistence of chaotic regions and invariant tori (the structures encountered in integrable systems).

We note also that the main phase space structures are dynamically independent since no trajectory
can cross from one region to the other.
We will show in section 4 that the very 
existence of these structures enables to organize the spectrum into independent frequency subsets.
In the following subsection, the generic character of these structures is checked by computing the PSS
at different rotation.


\subsection{Transition to chaos $\Omega \neq 0$}

The evolution of the dynamics with increased rotation is first described for $\tilde{L}_z=0$ and then for $\tilde{L}_z \neq  0$.

\subsubsection{The $\tilde{L}_z=0$ case}

The PSS computed at the three rotation rates $\Omega/\Omega_K=[0,0.15,0.32]$ respectively corresponding to the three flatness $\epsilon=[0,0.01,0.05]$ 
are displayed in Fig.~\ref{fig1} to illustrate the effect of a small centrifugal deformation on the ray dynamics.
This perturbation of the integrable $\Omega=0$ system is very similar to one described by the KAM-theorem
\citep{Chirikov,Leshouches,Ott,Gutzwiller,Lichtenberg,Lazutkin}.
Indeed, for a smooth small perturbation of an integrable Hamiltonian, this theorem states that the tori of the integrable system which survived
the perturbation occupy most of the phase space volume.
More specifically most of the irrational tori while being continuously perturbed can still be associated
with N invariants, thus keeping their torus structure. This is the case of the undulated lines
observed in the high $\tilde{k}_{\theta}$ domain of Fig.~\ref{fig1}. By contrast, all rational tori are immediately destroyed for a non-vanishing
perturbation. The KAM theorem implies that, despite the fact that the destroyed rational tori form a dense set in the phase space,
the volume they occupy
vanishes as the perturbation goes to zero.

\begin{figure}[t]
\resizebox{\hsize}{!}{\includegraphics{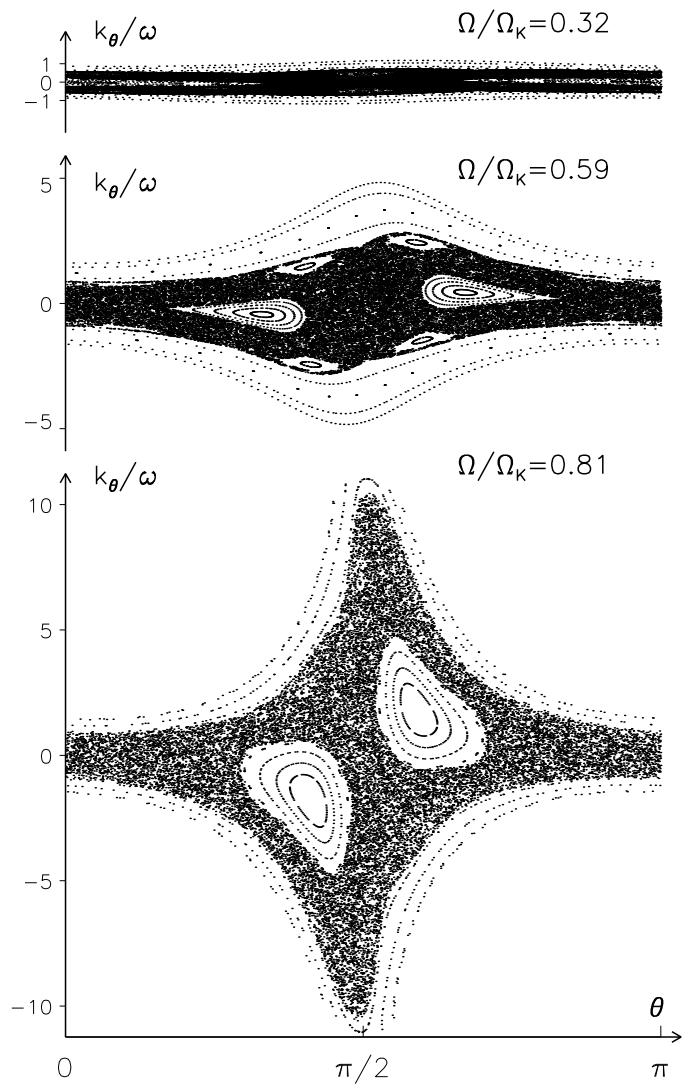}}
\caption{Three $\tilde{L}_z=0$ PSS visualizing the evolution of phase space as a function of rotation. All these PSS show
the 2-period periodic orbit island chain embedded in a central chaotic region. As rotation increases, the 2-period island chain
moves towards the equator while the central chaotic region enlarges. Note that the first $\Omega=0.32 \Omega_K$ PSS is
displayed with a different scale in Fig.~\ref{fig1}. The units are the same as in Fig.~\ref{fig1}.}
\label{fig2.lz0}
\end{figure}
\begin{figure}[t]
\resizebox{\hsize}{!}{\includegraphics{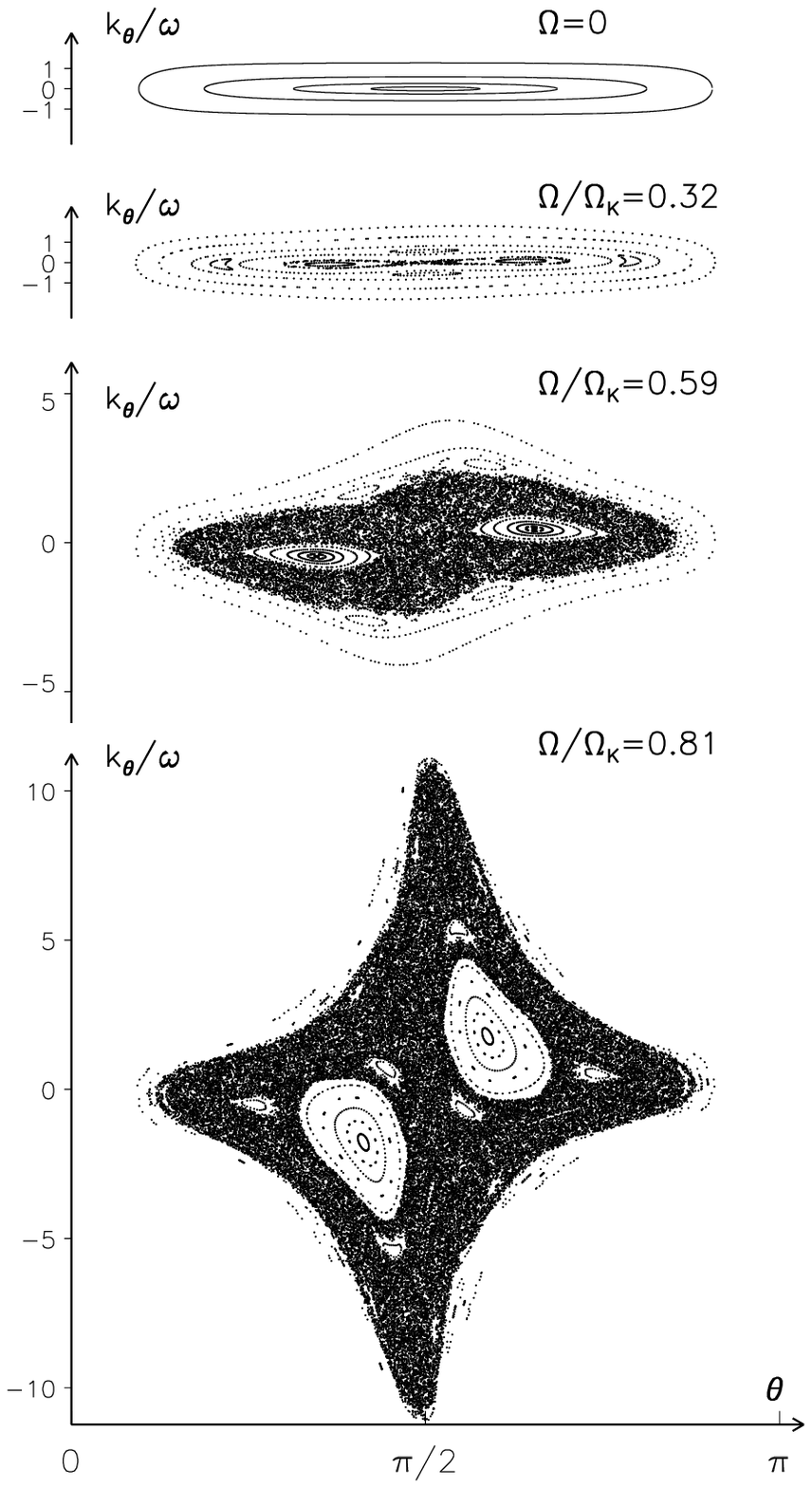}}
\caption{Three $\tilde{L}_z=0.4 /\omega_0$ PSS visualizing the evolution of phase space as a function of rotation. The evolution is qualitatively
 similar
to the $\tilde{L}_z=0$ case although the enlargement of the central chaotic region
occurs at higher rotation rates. Same units as in Fig.~\ref{fig1}.}
\label{fig2.lz5}
\end{figure}

The theory of KAM-type transition to chaos also describes how resonant tori disappear.
In our case, they correspond to one-dimensional curves on the PSS,
formed by families of periodic orbits.
All orbits of the same torus will have the same period $q$.
The so-called Poincar\'e-Birkhoff
theorem predicts
that a (smooth) small perturbation
will transform this one-dimensional curve into
a chain of $q$ stable points belonging to the same periodic orbit
and surrounded by stable islands, intertwined with $q$ unstable
periodic points. A small chaotic zone appears in the vicinity of the unstable
fixed points and grows with the perturbation.  The stable islands have themselves
an intricate self-similar structure of small island chains surrounded by invariant structure (tori).
This phenomenon is illustrated at $\Omega/\Omega_K =0.15$ in Fig.~\ref{fig1} where, near the $\tilde{k}_{\theta} =0$ curve, we can observe
the 2-period island chain around the $q=2$ stable periodic points
and the small chaotic region around the corresponding unstable points. This results from the destabilization of the 
resonant torus associated with the periodic orbits along the diameters of the spherical star.
Note that the width of the island chains (resonance width)
is expected to be approximately proportional to the square root of the perturbation, and decreases with $q$.

What occurs for large perturbations following the KAM-type transition of integrable Hamiltonians has been studied in many systems.
The general phenomenology which emerges also corresponds to what we observe
in our system
for increased rotation (see the PSS of Fig.~\ref{fig2.lz0} computed for 
$\Omega/\Omega_K=[0.32,0.59,0.81]$ corresponding to the flatness $\epsilon=[0.05,0.15,0.25]$).
The surviving irrational tori as well as the island chains are progressively destroyed. This leads to the enlargement of the chaotic regions which were
originally confined by these tori.
This is illustrated in Figs.~\ref{fig1} and ~\ref{fig2.lz0} where the surface of the central chaotic region becomes progressively larger with rotation.
The island chains typically undergo a series of bifurcations for increasing
perturbation; the most common bifurcation is the period-doubling one,
where a stable periodic orbit of period $q$ is changed to an unstable orbit
plus a stable orbit of period $2q$. As the sequence of bifurcations goes on,
stable orbits are of longer and longer period until they eventually disappear.
The destruction of stable regions is
visible between $\Omega=0.59 \Omega_K$ and  $\Omega=0.81 \Omega_K$
(Fig.~\ref{fig2.lz0}), as the 6-period island chain embedded in the chaotic central region at $\Omega=0.59 \Omega_K$ has
disappeared a higher rotation.
As mentioned above, the fact that the largest stable island originates from a short periodic orbit (here a 2-period periodic orbit)
is also a common feature of the KAM-type transitions to chaos.

While not visible on this figure, a zoom on other regions of the PSS would show the same process going on at small scales.
It is however clear that the irrational tori associated with large values of $\tilde{L}$ survive longer, this property
being also encountered in classical billiards \citep{Lazutkin}, where tori
close to the billiard boundary are the most robust.

\subsubsection{The $\tilde{L}_z \neq 0$ case}

Qualitatively, the transition to chaos is very similar to the $\tilde{L}_z =0$ case. This is shown
in Fig.~\ref{fig2.lz5} where PSS computed for $\tilde{L}_z = 0.4 / \omega_0 $, are shown for increased rotation.
The main effect of increasing $\tilde{L}_z$ is to delay the transition towards chaos to higher rotation rates.
Indeed by comparing PSS computed at the same rotation rate (see Fig.~\ref{fig2.lz0} and
Fig.~\ref{fig2.lz5}), one observes that the central chaotic region is 
more constrained by surviving tori for larger $\tilde{L}_z$ values. For example
at $\Omega=0.32 \Omega_K$ the central chaotic region is much more developed for $\tilde{L}_z=0$ than for $\tilde{L}_z=0.4 / \omega_0$.
The $\Omega=0.59 \Omega_K$ PSS provides another example since for $\tilde{L}_z=0.4 / \omega_0$ the island chain associated with the 6-period orbit
is separated by a surviving KAM tori from the central chaotic 
region while such a stable structure has already been destroyed for $\tilde{L}_z=0$. Finally, at $\Omega=0.81 \Omega_K$, we can observe that 
the central chaotic region for $\tilde{L}_z=0.4 /\omega_0$ contains visible surviving structures
while this is not the case for $\tilde{L}_z=0$.

The slower transition to chaos can be interpreted as being due to the angular constraint $-\arcsin(|\tilde{L}_z |/\tilde{L}) < \theta < \arcsin(|\tilde{L}_z |/\tilde{L})$
imposed on the dynamics. This is compatible with the fact that for infinite $\tilde{L}_z$ the
trajectories are confined to the equatorial plane and the dynamics becomes integrable because of the circular symmetry of this asymptotic limit.

\section{The asymptotic properties of the acoustic modes}

In this section, we show that ray dynamics provides a qualitative and
quantitative understanding of the high-frequency acoustic modes.

The question to be addressed is that of 
how to construct modes, that is standing waves, from the 
short-wavelength propagating waves described by ray dynamics.
Such modes construction is expected to be approximately valid in the asymptotic regime
of high-frequencies (this asymptotic regime is called the semi-classical regime in a quantum physics context).
As mentioned in the introduction, the answer depends on the nature of the Hamiltonian system. For integrable systems,
each phase space trajectory remains on an invariant torus and this enables to construct modes
from a positively interfering superposition of these travelling waves on the torus.
This is not any more
the case for chaotic systems, where the ray dynamics provides
no invariant structures on which to build modes.

Thus for integrable systems, the modes and the frequencies can in
principle be explicitly determined from the acoustic rays through
well-known formulas called Einstein-Brillouin-Keller (EBK) quantization after the name of its main contributors.
We shall recall the results obtained by \citet{Gough} when applying the EBK method to
spherical stars (sub-section 4.1).
While this procedure is not applicable to chaotic systems, 
other concepts and methods have been developed and tested in quantum physics 
to interpret the non-integrable dynamics. 
These concepts have been also applied to other wave phenomena such as those
observed in e.g. microwave resonators \citep{Stoc},
lasing cavities \citep{Nock},
quartz blocks \citep{Elle}
and
underwater waves \citep{Brow}.
Their potential interest for helioseismology has been suggested, although not demonstrated, by
\citet{Perd1} and \citet{Perd2}.
Here, we shall apply them to the non-integrable ray dynamics of rapidly 
rotating stars.
More specifically, we have seen in section 3 that the ray dynamics of such stars corresponds to a
mixed system where parts of phase space display integrable
behaviour and other parts chaotic dynamics.
In such case, the organization/classification
of modes in the semi-classical regime is expected to closely follow the structure of phase space.
Near-integrable regions of phase space like the island chains are amenable to EBK
quantization, leading to a regular frequency spectrum, while
the modes originating from chaotic regions have an irregular frequency
spectrum with generic statistical properties.
Another important information provided by ray dynamics is the averaged number of modes that can be constructed from a given
 phase
space region, this number being proportional
to the volume of the region considered.

In the following, we explicit these concepts and methods in the context of stellar
acoustics (sub-sections 4.1, 4.2, 4.3).
Then, their relevance to describe the asymptotic properties of the acoustic modes is tested 
by confronting their predictions to the actual
properties of
high-frequency
acoustic modes (sub-sections 4.4, 4.5, 4.6, 4.7).
These modes are axisymmetric modes in the 
frequency range $[9 {\omega}_1 , 12 {\omega}_1 ]$, $\omega_1$ being the lowest acoustic mode frequency of the stellar model considered.
They have been computed for a $\Omega = 0.59 \Omega_K$ uniformly rotating $\mu=3$ polytropic model of star and under the same assumptions
as for ray dynamics (adiabatic perturbations, no Coriolis acceleration, Cowling approximation).

\subsection{The integrable case $\Omega=0$}

To build modes from the ray dynamics, the 
wave-like solution $\hat{\Psi} = A({\bf x}) \exp [i \Phi ({\bf x})] $
is regarded as a function of the phase space variables $\vec{x}, \vec{k} = \nab \Phi$ that is subsequently projected onto the position space.
The condition that  $\exp [i \Phi ({\bf x})]$ be single-valued on the position space requires that for any phase space trajectory that closes on itself 
in the position
space, the variation of $\Phi$ along this closed contour is a multiple of $2 \pi$. As trajectories of an integrable system stay on invariant tori,
this condition leads to the EBK formula:
\begin{equation} \label{quant}
\int_{C_j} \vec{k} \cdot \vec{dx} = 2 \pi (n_j + \frac{\beta_j}{4} )
\end{equation}
where $C_j$ is any closed contour on a given torus and $n_j$ and $\beta_j$ are non-negative integers.
The integer $\beta_i$ called the Maslov index is introduced to account for a $\pi/2$ phase lag that must added
each time the contour crosses
a caustic. Indeed, the caustic corresponds to the boundary of the torus projection onto position space;
the amplitude $A$ taken in the position space is discontinuous there, leading to the $\pi/2$ phase loss (see \citet{Keller} for details).

While this expression is valid for any closed contour on the torus, it can be shown
that it gives the same condition for all
contours that can be continuously deformed to the same one.
Thus
in fact the EBK quantization yields $N$ independent conditions, as only $N$
topologically independent closed paths exist on a $N$-dimensional torus.  As these closed paths
do not need to
be actual trajectories of the dynamical system, 
the
usual way to construct EBK solutions is to choose contours for which the formulas are simple to compute.
The quantization conditions thus
select a particular torus on which a mode can be
built, irrespective of the fact that the torus is resonant or non-resonant.
For spherical stars, three independent contours on a torus specified by $L$, $L_z$ and $\omega$ can be obtained by varying one of the
spherical coordinates and fixing the other two. Using similar contours, \citet{Gough} obtained the three conditions:
\begin{eqnarray} \label{quant2}
\int_{r_i}^{r_e} \!\left( \frac{\omega^2 -\omega_c^2}{c_s^2} - \frac{L^2}{r^2} \right)^{1/2} \!\!\!\!\!dr = (n-\frac{1}{2}) \pi,  & \; L=\ell+ \frac{1}{2}, & \; L_z = m
\end{eqnarray}
\noi where $n$, $\ell$, $m$ are non negative integer, $r_i$ and $r_e$ being the internal and external caustic respectively. Note that $L=\omega \tilde{L}$
and $L_z=\omega \tilde{L}_z$. 
The associated eigenmodes have been also explicitly constructed
from the trajectories on the selected torus.
As shown by \citet{Gough}, the result of this EBK quantization is practically identical to the usual asymptotic theory 
of acoustic modes in spherical stars
which uses the separability of the three-dimensional eigenvalue problem and a WKB approximation of the resulting 1-D boundary value problem in the radial
direction (in the usual analysis, $L= \sqrt{\ell(\ell+1)}$, differs from the EBK result, especially at low $\ell$ values).
This shows that the integers $n,\ell,m$ derived from the EBK quantization conditions
do correspond to the order, degree and the azimuthal number of the acoustic modes in spherical stars.

The tori on which the eigenmodes are constructed can be visualized on the PSS.
For example, the $(\ell,n,m)=(8,10,0)$ mode is associated
with the torus $\tilde{L} = \pm (\ell +1/2)/\omega_{n,\ell,m}$, $\tilde{L}_z=0$ which imprint on the PSS are the straight lines $\tilde{k}_{\theta} = \pm \tilde{L}$.
The intersection of various mode-carrying tori with the $\tilde{L}_z=0$ PSS are shown in Fig.~\ref{fig1}. High radial order modes approach the central
$\tilde{k}_{\theta} =0$ line while large degree modes 
occupy the high $\tilde{k}_{\theta}$ region.

\subsection{Chaotic systems}

It has been widely recognized
in the past decades that most dynamical systems are not integrable
and therefore display
various degrees of chaos. The quantum chaos field has
studied quantum systems whose short-wavelength classical limit displays such
chaotic behavior. As has been recognized early by
\citet{Einstein} the EBK quantization explained in the above
paragraph cannot be applied to such systems. Indeed, no
$N$-dimensional invariant structure exists on which to apply conditions of constructive interference like the EBK condition.
Rather, the semi-classical limit
of these chaotic systems yields a Fourier-like formula
which connects the set of all classical periodic orbits
to the whole spectrum.  This formula, called Gutzwiller trace
formula \citep{Gutzwiller} is much more delicate
to use than EBK formulas, since it represents a divergent sum
from which information can be extracted only through refined
mathematical and numerical methods.

On the other hand, the very complexity of chaotic systems leads to
statistical universalities. Indeed, in a similar way as statistical
physics emerges from the random behavior of individual particles,
in chaotic systems the randomness induced by chaos leads to robust
statistical properties of eigenmodes.  Contrary to modes
of integrable systems, which are localized on individual tori selected
by the EBK conditions, in chaotic systems modes are generally
not associated to a specific structure in phase space, and
are ergodic on the energy surface.
It has been found that one can model such systems
 by replacing the Hamiltonian by a matrix whose
entries are random variables with Gaussian distributions.
Such ensembles of {\it Random Matrices}, which contain no free parameter
but take into account the symmetries of the system,
can give precise predictions, which have been
found to describe accurately many statistical properties
of the modes of systems with a chaotic classical limit.
This has been conjectured and checked numerically for many systems
\citep{Bohib,Leshouches}.

The comparison with Random Matrix predictions
is often done through the statistical analysis of the frequency spectrum.
In general
the density of modes as a function of the frequency
\begin{equation}
d(\omega)=\sum_n \delta(\omega-\omega_n)
\end{equation}
where $\delta$ is the delta function, can be written as the sum of two functions
\begin{equation} \label{dofE}
d(\omega)=d^{av}(\omega)+ d^{fluct}(\omega).
\end{equation}
The quantity $d^{av}(\omega)$ (hereafter called the Weyl term) is a smooth function which
describes the average density of modes at a given frequency. It has been
known from the beginning of the twentieth century that this term can
be calculated 
from general geometric features of the system such as
phase space volume and therefore is independent of the chaotic or integrable nature of the dynamics
($d^{av}(\omega)$ is estimated for stellar acoustic modes in subsection 4.7).
In contrast, the function $d^{fluct}$ which describes the fluctuations
around the mean position of eigenfrequencies
strongly depends on the nature of the dynamics.
(Note that most textbooks on this subject use the quantum physics terminology, that is 'energy level' instead
of frequency and 'density of states' instead of density of modes).

The spectra of integrable systems are
predicted to be uncorrelated, and in general this leads to fluctuations
given by the Poisson distribution \citep{tabor}. In contrast,
for chaotic systems these fluctuations should be given by Random Matrix Theory.
 The theory of Random Matrices
has therefore been developed in order to compute analytically
the predictions for specific quantities, which in turn could be compared
to numerical data for real systems.  A popular quantity to describe
fluctuations in spectra is the {\em spacing distribution $P(\delta)$}, which
is the distribution of spacings in frequency between consecutive eigenfrequencies,
once the frequency differences have been rescaled by
the Weyl term such that the average spacing is one.
The function $P(\delta)$ measures the correlations at
short
distances in frequency in the spectrum, and therefore does not give
information
about all statistical properties, but is nevertheless very useful
since the predictions are
strikingly different for the Poisson distribution and for Random
Matrix Theory.  While the Poisson distribution corresponds to
$P(\delta)=\exp(-\delta)$, the prediction of Random Matrix Theory is the Wigner
distribution $P(\delta)=\pi \delta/2 \exp(-\pi \delta^2/4)$, which displays frequency
repulsion (level repulsion in the quantum terminology) at short
distances (small $\delta$) and falls off faster than Poisson at large $\delta$.

\subsection{Mixed systems}

We have seen
in section 3 that the acoustic ray dynamics in rotating stars has a mixed 
character as chaotic regions coexist with stables structures like island chains or invariant tori.
Such mixed system are actually the most common in nature, completely integrable and chaotic
systems being limiting cases.

In the context of quantum chaos, seminal studies of these systems 
by \citet{percival} and \citet{robnik}
enabled to conjecture that a good description of their spectrum at high
energy
is obtained by quantizing independently the structured and chaotic
parts. While a zoom on island regions would reveal a complex structure involving chaotic trajectories and chains of smaller islands, these
small scale details can be neglected for the island quantization if the mode wavelength remains large as compared to these scales.
Instead, the presence of a large number
of invariant structures constrains enough the dynamics to make
the system similar to a purely integrable structure to which EBK quantization
applies. These region are then called near-integrable.

Thus subsets of modes can be associated to the different
near-integrable island chains, while other subsets correspond to the chaotic zones.
In each subset, the modes behave as if they were constructed from
an isolated system.
Thus in mixed systems the frequency spectrum
can be described as a
superposition of independent frequency subsets associated with the
different phase space regions.  Subsequent works have shown this
picture to be a good approximation of actual spectra, although in
some cases certain correlations are present between the frequency
subsets due to the presence of partial barriers in phase space or
to the existence of modes localized at the border between zones
\citep{Bohia}.

The acoustic ray dynamics of rapidly rotating stars being
of this mixed type, one can expect such an organization of the
spectrum to be valid, even though the modes are of quite large
wavelength
compared to previous studies in quantum chaos.
To test this hypothesis, it is convenient to compute a
phase-space representation of the modes.  Indeed, the chaotic
or near-integrable zones are well-defined in phase space while their projections
in position space, were modes are usually pictured,
 are generally much more difficult to separate.

\subsection{Associating modes to rays}

Constructing phase space representations for
modes has been envisioned since the beginning
of quantum mechanics, since it enables to test the
quantum-classical correspondence accurately.  
Contrary to states of a classical system which are defined by a point in phase space, modes have always a finite extension in phase space
since they have a finite wavelength and their localization in wavenumber space
is, according to Fourier
analysis, inversely
proportional to their localization in physical space.
Any mode occupies a finite volume of the order of $(2\pi)^N$ in the physical/wavenumber phase space
(a $(2\pi/\omega)^N$ volume in the physical/scaled-wavenumber ($\vec{x}$, $\vec{\tilde{k}}$) phase space or
a $(2\pi \hbar)^N$ volume in the position/momentum phase space of quantum physics).
\citet{wigner} was the first to construct a phase space function representing
a mode, but this so-called Wigner function
has the disadvantage of being positive or negative, and
therefore cannot be interpreted as a probability distribution of the
mode.
To circumvent this problem, one way is to smooth the Wigner
function by a Gaussian convolution.  The resulting function, called
Husimi distribution function \citep{husimi}, is always real and nonnegative and
can be understood equivalently
as the projection of the mode onto
a Gaussian wave packet centered around $\vec{x}$ and  $\vec{\tilde{k}}$:
\begin{equation}\label{hus1}
{\cal H}(\vec{x},\vec{\tilde{k}})= \left| \int \Psi(\vec{x}') \exp (-\|\vec{x}'-\vec{x}
\|^2/(2\Delta_x ^2)) \exp (i \omega \vec{\tilde{k}} \cdot \vec{x}')
d\vec{x}' \right| ^2
\end{equation}
\noi where $\Psi(\vec{x}')$ is the mode and $\exp (-\|\vec{x}'-\vec{x}
\|^2/(2\Delta_x ^2)) \exp (i \omega \vec{\tilde{k}})$ the Gaussian
wavepacket. 
In this expression,
the width
of the wavepacket $\Delta_x$ determines the resolution of the Husimi function in the spatial direction, the resolution in the scaled-wavenumber 
$\Delta_{\tilde{k}}$ being such that $\Delta_x \Delta_{\tilde{k}} \approx 1
/\omega$. These quantities determine the minimal extent of the mode representation in both directions.

The computed modes are 3D modes and they shall be compared with the reduced ray dynamics computed
on a 2D meridional plane. As shown in the spherical case by \citet{Gough}, the amplitude of a 3D axisymmetric mode constructed 
from acoustic rays obtained on neighboring
meridional planes decreases as $(r \sin \theta)^{(-1/2)}$ because away from the rotation
axis the distance between the planes and thus
the density of rays increases.
Thus, the computed 3D modes have been scaled by $(r \sin \theta)^{(1/2)}$ to better
represent the mode amplitude on a meridional plane.
Moreover to obtain a phase-space representation limited to the PSS, we
actually computed the Husimi's distribution function of the
1D
cut of the mode taken along the PSS:
\begin{equation}\label{hus2}
{\cal H}(s,\tilde{k})= \left|\int \Psi'(s') \exp (-(s'-s)^2/(2\Delta_s^2)) \exp (i\omega
\tilde{k} s') ds'\right|^2
\end{equation}
\noi where $\Psi'$ is the scaled version of the mode $\hat{\Psi}=\hat{P}/\alpha$ solution of Eqs.~\eq{div}, \eq{vel} and \eq{adia},
$s$ is a curvilinear coordinate
along the curve $r=r_p$ and $\tilde{k} = \vec{\tilde{k}} \cdot \vec{e}_p$ is
the scaled-wavenumber in the direction tangent
to this curve, $\vec{e}_p$ being a unit vector tangent to the curve. ${\cal H}$ is then expressed in terms of the PSS coordinates
using the following relations between $[\theta, \tilde{k}_{\theta}]$ and $[s,\tilde{k}]$:
\begin{eqnarray} \label{hus3}
s=\int_{\theta_0}^{\theta} \left({r}_p^2 + \left( \frac{d r_p}{d \theta}\right)^2 \right)^{1/2} \!\!\!\! d\theta' & \;\tilde{k} =
\tilde{k}_{\theta} \vec{E}^{\theta} \cdot \vec{e}_p
\end{eqnarray}
The vector $\vec{E}^{\theta}$ is defined in appendix B.2. The integral \eq{hus2} is performed in the interval $[\theta-\pi, \theta+\pi]$, the
mode being prolonged by symmetry outside the $[0, \pi]$ interval.

\begin{figure}
\resizebox{\hsize}{!}{\includegraphics{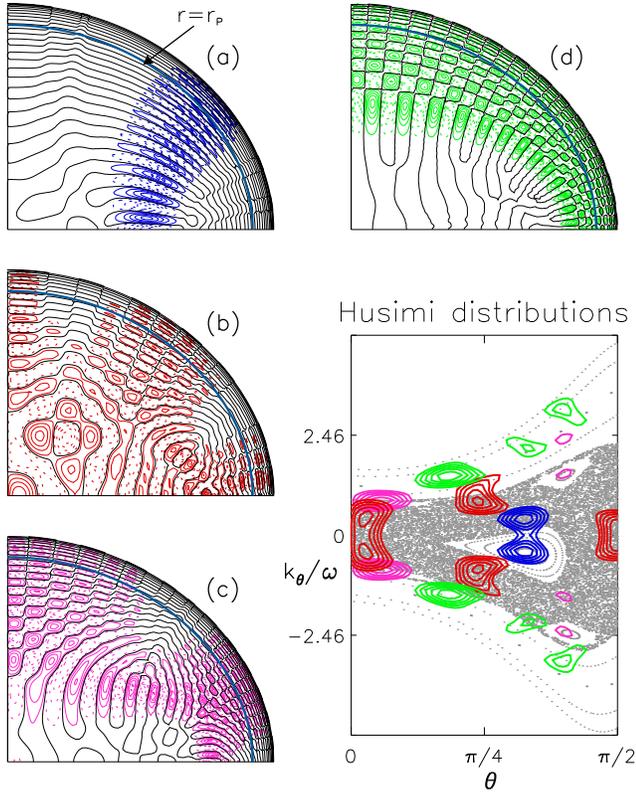}}
\caption{(Color online) Four axisymmetric modes and their phase space representation : (a) a 2-period island mode (blue/dark grey), 
(b) a chaotic mode (red/grey), (c) a 6-period island mode (magenta/light grey) and (d) a whispering gallery
mode (green/light grey). The amplitude distribution of the
scaled mode $\Psi'$ is represented by its nodal lines (black) and its 
positive (full colored/grey lines) and negative (dashed colored/grey lines)
level curves.
The Husimi distributions of the four modes computed for the same $\Delta_s = 0.12 R_e$ are represented by their level curves.}
\label{fig.hus}
\end{figure}

The Husimi function has been computed for the axi-symmetric modes of the $\Omega = 0.59 \Omega_K$ star
and its contour plot is compared with the PSS of the ray dynamics computed for the same star model.
Fig.~\ref{fig.hus} illustrates this process by showing the position space as well as the phase space
representation of four typical modes.
As can be observed, the modes can be clearly associated with one of the main
structure of the phase space namely, the 2-period island chain, the large central chaotic region,
the 6-period island chain or the whispering gallery region.
On the PSS, we note however that the Husimi function is symmetric with respect to $\tilde{k}_{\theta}$ while the dynamics is not.
This difference is due to the fact that the PSS is constructed only with $\tilde{k}_r >0$ intersecting trajectories
while the Husimi function computed from the
mode cut on the $r=r_p$ contains no information about the sign of $\tilde{k}_r$.
Nevertheless, in the high-frequency interval $[9 \omega_1, 12 \omega_1]$ that we studied in detail,
any ambiguity on the phase space location can always be resolved using the additional information
on the mode distribution in the position space. 
In this frequency interval, we thus classified the modes according to their localization in phase space
distinguishing the 2-period island modes, the chaotic modes, the 6-period island modes
and the whispering gallery modes associated with the corresponding phase space regions.
As a result, the full frequency spectrum can be discomposed into sub-spectra associated with phase space structures.
Fig.~\ref{fig.spec} displays the four sub-spectra. 

In the following, 
we shall analyze these sub-spectra and test whether
the Percival and Berry-Robnik conjecture described in subsection 4.3 applies to acoustic modes in rapidly rotating stars.
We first study the regular character of the sub-spectra issued from near-integrable phase space regions (sub-section 4.5)
and then consider the spectrum of chaotic modes (sub-section 4.6).

\subsection{The regular spectra}

\begin{figure}
\resizebox{\hsize}{!}{\includegraphics{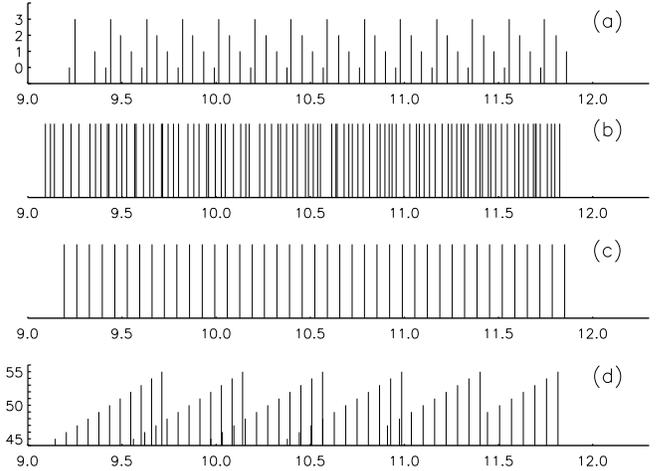}}
\caption{Frequency sub-spectra of four classes of axisymmetric modes : (a)
the 2-period island modes, (b) the chaotic modes antisymmetric with respect to the equator, (c) 
the 6-period island modes and (d) some whispering gallery modes. For the sub-spectra (a) and (d), 
the height of the vertical bar specifies one of the two quantum numbers characterizing the mode.}
\label{fig.spec}
\end{figure}

A spectrum is said to be regular if it can
be described by a function of $N$ integers, $N$ being the degree of freedom of the system.
In accordance with previous studies by \citet{Lign}, \citet{pre} and \citet{Rees8}, the spectrum 
of the 2-period island modes 
is well-fitted by the empirical
formula
\begin{equation}
\omega_{n \ell} = n {\delta}_n + {\ell} {\delta}_{\ell} + \alpha
\label{numbers}
\end{equation}
\noi confirming the regular nature of this spectrum that is also clearly apparent on Fig.~\ref{fig.spec}(a).

The 6-period island mode spectrum shown in Fig.~\ref{fig.spec}(c) is also regular as it is closely fitted by the even simpler formula:
\begin{equation}
{\omega}_{n'} = n' {\delta}_n' + \alpha'
\label{numberss}
\end{equation}
\noi Indeed, the root mean square error between this empirical fit and the actual spectrum is equal to 1.9 percent of ${\delta}_n'$
(where ${\delta}_n'$ has been determined as the mean of the spacing between
consecutive frequencies and $\alpha'$ is fixed such that the model is exact at a reference frequency).

While a simple linear law such as Eqs.~\eq{numbers} or \eq{numberss} does not apply to the whispering gallery modes, there are strong
evidences that this sub-spectrum is also regular. Thanks to the regularity of the nodal lines pattern (as apparent
in Fig.~\ref{fig.hus}(d)), two integers 
corresponding to the number of nodes along the polar axis
and to the number of nodes following the internal caustic can be easily attributed to each modes.
When plotted against the number of caustic nodes (as in Fig.~\ref{fig.spec}(d)), the spectrum 
clearly shows a regular behavior. The fact that the function of these two integers describing the spectrum is however not
as simple as Eqs.~\eq{numbers} or \eq{numberss} is expected from what we know about the
regular spectrum of high degree modes in spherical stars (see for example \citet{Chris80}). 

The regularity of the three sub-spectra issued from near-integrable phase space regions 
is fully in accordance with the Percival's conjecture.
An important consequence is that theoretical model of these spectra can in principle
be obtained from the EBK quantization
of the invariant structures of the corresponding near-integrable regions.
As a result we should be able to
relate the potentially
observable quantities ${\delta}_n$, ${\delta}_{\ell}$ or ${\delta}_n'$ 
to the star properties. 
In practice, the standard method is first to construct
the normal forms around the central periodic orbit in order to describe
the dynamics in the island, and then use the EBK
quantization scheme to find the asymptotic formula for the modes
\citep{Bohia, Lazutkin}.
While such a program is outside the scope of the present paper, we mention below the result
obtained in \citet{pre} for the 2-period island modes using an
equivalent procedure, which may be more physically transparent, and extend it to the 6-period island modes.

As already noted, the propagation of acoustic waves in
our system is similar to the propagation of light in an inhomogeneous
medium, the role of the medium index being played by $1/c_s$. The construction of standing wave
solutions between two bounding surfaces has
been investigated intensively in the context
of the study of laser modes in cavities \citep{Koge} and it consists
in applying the paraxial approximation in the vicinity
of the optical axis. While generally applied to homogeneous media, this approximation can be extended
to the non-homogeneous case as in \citet{Perm}. Applying this formalism to the acoustic modes
associated with periodic orbits,
\citet{pre} found a model spectrum equivalent to Eq.~\eq{numbers} with
\beqan
{\delta}_n= \frac{\pi}{\int_{\mbox{a}}^{\mbox{b}} d\sigma/c_s}  &\;\;\mbox{and}\;\;  & {\delta}_{\ell}=
2 \frac{\int_{\mbox{a}}^{\mbox{b}} c_s d\sigma /w^2}{\int_{\mbox{a}}^{\mbox{b}} d\sigma/c_s}
\eeqan{deltas}
\noi where $\sigma$ is the curvilinear coordinate along the periodic orbit and the integral is computed between
the end points of the orbit (these points are shown in Fig.~\ref{fig3} for the 2-period and the 6-period periodic orbits
being denoted (a,b) and (a',b') respectively). 
The quantity $w(\sigma)$ in the expression of ${\delta}_{\ell}$ describes the 
spreading of the wave beam in the direction transverse to the periodic orbit and verifies a differential equation
which depends on the sound speed and its
transverse derivative taken along the periodic orbit.
Moreover the integers $n$ and $\ell$ correspond respectively to the number of nodes in the directions parallel and transverse
to the orbit (see \citet{pre} for details).

When applied to the 2-period periodic orbit, this theoretical expression of 
${\delta}_n$ yields a value  $0.5635 \omega_0$ which differs from its empirical value 
by only 2.2 percent. 

The 6-period island mode spectrum can be modeled in the same way. In the frequency 
interval considered, these modes
have a similar structure in the direction transverse to the central orbit and should therefore be 
associated with the same $\ell$ value.
The model spectrum has thus the same form as Eq~\eq{numberss} where
\beqan
{\delta'}_n= \frac{\pi}{\int_{\mbox{a'}}^{\mbox{b'}} d\sigma/c_s} 
\eeqan{deltab}
\noi 
As for the 2-period modes, we find that this theoretical value of ${\delta'}_n$ differs by only a few percents from the
empirical determination of ${\delta'}_n = 0.186 \omega_0$.

These two examples show that ray dynamics can provide a quite accurate model of the near-integrable spectra
in the relatively low frequency regime considered. Moreover 
the model \eq{numbers} of the 2-period island mode spectrum remains reasonably 
accurate at lower frequencies \citep{Lign} and can be extended to non-axisymmetric modes \citep{Rees8}.

\subsection{The chaotic modes}

\begin{figure}
\resizebox{\hsize}{4.cm}{\includegraphics{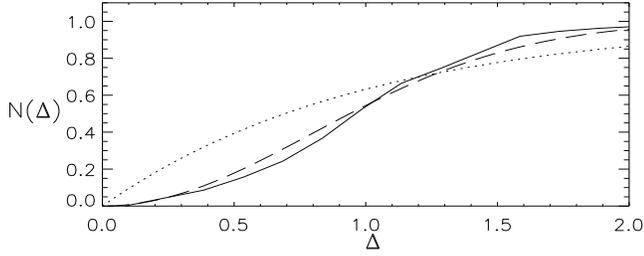}}
\caption{
Integrated spacing distribution $N(\Delta)$ of the chaotic modes 
(full line). 
The dashed line shows $N(\Delta)$ for the Gaussian Orthogonal Ensemble of Random Matrix Theory while
the dotted line shows the integrated Poisson
distribution typical of integrable systems.}
\label{fig.int}
\end{figure}

A large subset of the frequencies correspond
to modes localized in the chaotic zone of phase space (the chaotic modes).  As we have seen
in subsection 4.2, one should not expect regular patterns for this part of the
spectrum.  Rather, the chaotic character of the phase space should be reflected
in specific statistical properties of the sub-spectrum, which should follow
predictions from Random Matrix Theory.  To test this predictions, we
have studied the distribution of the consecutive frequency spacings
$\delta_i = \omega_{i+1} - \omega_{i}$ of the chaotic modes.
Fig.~\ref{fig.int} shows the integrated
distribution $N(\Delta)=\int_0^{\Delta} P(\delta) d\delta$ (with spacings normalized by the mean
level spacing within the chaotic subset, as should be done).
The distribution is constructed from the two independent distributions obtained for
the equatorially symmetric and anti-symmetric modes, corresponding to around $187$ modes in total.
Although the difficulty to solve Eqs.~\eq{div}, \eq{vel}, \eq{adia} prevents us to reach as
large frequency samples as can be obtained for other systems \citep{Bohib},
the numerically computed $N(\Delta)$ agrees well with the Random Matrix predictions,
and is clearly different from the Poisson distribution typical of
integrable systems.  This result indicates that these modes, selected
by means of the comparison between their localization in phase space and
the ray dynamics, have indeed the frequency statistics expected for chaotic
modes.
We therefore think that this confirms the validity of the ray model,
and gives a strong evidence that wave chaos actually
occurs in rapidly rotating stars.

We note that the modes we identify as chaotic are located in the chaotic zone
but cover only part of it, as for the example of Fig.~\ref{fig.hus}.  We think
this is partly due to the relatively low frequency considered, which prevents
ergodicity of the modes to be clearly visible.  In addition, it is known
that certain low-energy eigenfunctions of chaotic systems called scars are concentrated along short
periodic orbits of the system \citep{scars}.
In this case, rather than being ergodic,
some individual modes are effectively localized in the vicinity
of such orbits.  
This effect can
create some subsequences of low-energy modes
with regular frequency patterns, even if the distribution of frequency spacings
follows the predictions of Random Matrix Theory.  Precise investigation
of this phenomenon in the context of stellar acoustic modes may be
important.

\subsection{Predicting the number of modes in each sub-spectrum : The Weyl formula}

In this subsection, we show that ray dynamics enables to estimate the number
of chaotic and island modes present within a given frequency interval. This information
is complementary to the regular/irregular properties of the associated sub-spectra shown in the
previous section and it is crucial to build an asymptotic model of the frequency spectrum.

We have seen in subsection 4.2 that the density of modes as a function of the frequency
$d(\omega)$ can be written as the sum of a smooth part $d^{av}(\omega)$ and
an oscillatory part $d^{fluct}(\omega)$ (see Eq.~\eq{dofE}).  
Weyl, at the beginning of the twentieth century, derived analytically an asymptotic expansion of $d^{av}(\omega)$ \citep{Wey}.
The leading term of the Weyl formula can be obtained from
general principles.
We have seen (subsection 4.4) that in average a mode occupies a $(2\pi)^N$ volume
in the physical/wavenumber phase space.
The averaged number
of modes in a given phase space volume can thus be estimated as the volume of phase space
available divided by $(2\pi)^N$, the volume occupied by one mode.
In the following, we shall first verify that the leading term of the Weyl formula 
yields a reasonable estimate of the number of modes in the case of spherically symmetric stars.
Then, we shall calculate the phase space volume of the chaotic and island regions
and confront the result with those obtained from the numerical computation of modes at $\Omega = 0.59 \Omega_K$.

The Hamiltonian formulation $H'=\omega=\sqrt{c_s^2\vec{k}^2 + \omega_c^2}$ is best suited
for this calculation. The averaged
number of modes below a given frequency $\omega$ reads:
\beq \label{eq:weyl}
\bar{N}(\omega)=\frac{1}{(2\pi)^N}{\cal V}(\omega)
\eeq
where ${\cal V}(\omega)$ is the phase space volume corresponding to energies smaller than $\omega$ defined as
\beq \label{eq:volume}
{\cal V}(\omega)=\int_{{\cal A}(\omega)} d \vec{k}^N d \vec{x}^N
\eeq
\noi where ${\cal A}(\omega)$ is the phase space region defined as $H'(\vec{x},\vec{k}) \le \omega$.

For spherical stars, the dynamics can be reduced to the 
one-degree-of-freedom dynamics characterized by the reduced Hamiltonian
$H^{sph} = \sqrt{c_s(r)^2(k_r^2 + L^2/r^2)+ \omega_c^2}$.
Applying the above formula to $H^{sph}$, the double integral ${\cal V}(\omega)$
can be integrated over $k_r$ to give:
\beq
\bar{N}^{sph}(\omega)=\frac{1}{\pi} \int_{r_i}^{r_e} \sqrt{\frac{\omega^2 - \omega_c^2}{c_s^2} - \frac{L^2}{r^2}} dr
\eeq
\noi which is thus the estimated number
of modes of given $L$ and $L_z$ with a frequency smaller than $\omega$.
This estimation can be compared to the results of the usual asymptotic theory (see Eq.~\eq{quant2}). Accordingly,
$\bar{N}^{sph}(\omega) = n-1/2$, where $n$ is the radial order associated to the frequency $\omega$ but is also the 
number of modes below the frequency $\omega$ for fixed values of $L$ and $L_z$.
We therefore conclude that in the 1D-spherical case the first term of Weyl's formula
yields a reasonable approximation of the averaged mode density.

In rotating stars, to estimate $\bar{N}(\omega)$ the number of modes below $\omega$ for a given $L_z$, we use 
the two-degree-of-freedom Hamiltonian $H^{'}_r = \omega=\sqrt{c_s^2 (\vec{k}_p^2 + L_z^2 /(r \sin \theta)^2 ) + \omega_c^2}$,  
and integrate the 4-dimensional volume integral in the wavenumber directions to obtain:
\beq
\bar{N}(\omega)=\frac{1}{4\pi} \int \!\!\!\!\! \int_{S_m} \frac{\omega^2 - \omega_c^2}{c_s^2} - \frac{L_z^2}{(r \sin \theta)^2} dS_m
\eeq
\noi where $S_m$ is the portion of the meridional plane surface where the integrand is positive.
While providing the total number of modes, this expression does not give the fraction of chaotic, island or whispering gallery modes
which are important quantities to model the spectrum.
These quantities can nevertheless be obtained by computing the corresponding phase space volumes and by applying Eq.~\eq{eq:weyl}.
This has been done for $\Omega/\Omega_K=0.59$ since at this rotation rate we can compare the results of Weyl's formula
to the numbers of chaotic and island modes obtained from the direct mode computation and the 
classification through the Husimi distribution.

The 4-dimensional phase space volumes have been evaluated using a Monte-Carlo quadrature method: points are randomly
chosen in a known volume $V_M$ that includes the volume $V$ to be computed. The proportion of points
inside $V$ approximates the ratio $V/ V_M$, thus providing an approximate value of $V$.
To decide
whether a given point
is inside or outside $V$, we used
space filling trajectories on the torus delimiting the volume $V$.
Two phase space volumes have been computed at $\Omega/\Omega_K=0.59$. The first one 
includes the large chaotic region as well as the island chains around the 2-period and 6-period orbits.
The second volume corresponds to the 2-period island chain.
The details of the calculation and an estimation of the error on the volume determination are given in Appendix C.

As a result, the leading term of the Weyl formula yields $34 \pm 2$  modes in the 2-period island chain
and $270 \pm 8$ modes outside the whispering gallery region in the frequency interval $[9.42 \omega_1, 11.85 \omega_1]$.
This value has to be compared with the $50$ island modes and the $276$ modes outside the whispering gallery region
obtained using the Husimi phase space representation of the modes computed in the same frequency interval.

The difference between the estimation given by \eq{eq:weyl} and the actual
number of modes in each subset of the frequency spectrum corresponds
most likely to the next order in the asymptotic expansion of the density
of modes.  Indeed, Eq.~\eq{eq:weyl} is only the first term in an asymptotic
expansion, the next term being usually proportional to the length
of the boundary between phase space zones.  At relatively low frequency,
this term can be significant.  Another source of imprecision
can stem from the fact that at such relatively low frequency some
partial barriers in phase space can trap island-like modes in the
vicinity of the island, enlarging the effective size of the regular
zone. Indeed, 
for some of the modes classified 
as island modes, the Husimi distribution is not entirely inside the island, the outer part remaining
close to the island.
Nevertheless, our study show that Eq.~\eq{eq:weyl} gives a reliable
estimate of the relative sizes of the frequency subsets which can be
obtained without any knowledge of the spectrum itself.

\section{Discussion and applications to asteroseismology}

In this section we discuss the validity of the assumptions
of our asymptotic analysis, and the implications of our results for the asteroseismology of rapidly rotating stars.

\subsection{Assumptions of the asymptotic analysis}

The WKB assumption underlying the asymptotic analysis is not justified for the lowest frequency acoustic modes.
While determining the limit of validity of the different results presented here is outside the scope of the paper,
we know that the regularities of the 2-period island mode sub-spectrum are relevant
down to about the $5^{\mbox{th}}$ radial order acoustic pulsations
(see \citet{Lign} and \citet{Rees8} for details).

Another important precaution when applying the present analysis stems from the fact that
the interpretation of the ray dynamics depends on the frequency range considered.
Indeed, if extremely high frequency modes were to exist,
their properties would closely follow the phase space structure up to its smallest details.
This means for example that such modes could be associated with the very small chaotic regions
which exist inside the island chains or in between the surviving KAM tori of the whispering gallery regions.
On the other hand,
finite-wavelength effects have to be taken into account when interpreting the ray dynamics at finite frequencies.
For example, the regularities of the island mode spectra in the $[9 \omega_1, 12 \omega_1]$ interval show
that the small chaotic zones within the island chain can be overlooked in this frequency range.
The same reasoning holds if one wants to interpret the ray dynamics at small rotation rates (see Fig.~\ref{fig1}).
The tiny chaotic regions predicted
by the KAM theorem could be interpreted as a proof for the existence of a chaotic mode frequency subset at vanishingly small rotation.
However, these modes should have such a short wavelength to "fit in" the chaotic region
that they may simply not exist (because they are strongly dissipated by diffusive effects
or their frequency is so high that they are not reflected at the surface).

Ray dynamics can not account directly for the coupling effects between two modes associated with two dynamically isolated
regions of phase space (as occurs for example in the well known tunneling effect). Indeed, while trajectories can not cross the dynamical barrier between
the chaotic and the island chain regions,
an island mode can be present on either side of the barrier if its frequency is very close
to the one of a chaotic mode (and vice versa). As usual for mode avoided crossings, the mode distribution can thus
be significantly perturbed by the coupling
but the frequency is only slightly affected especially in the high-frequency regime.
Quantifying the effects of such avoided crossings would require a specific study.

The WKB assumption that the wavelength is much shorter than the typical background lengthscale
breaks down for low-frequency acoustic mode but also when the typical lengthscale
of the stellar model becomes very small. This can occur in real stars, especially at the upper limit
of the core convective zone where strong composition gradients built in during evolution.
The effect of such a discontinuity has been studied for spherical stars \citep{Vor,Gou90,Ballot} and has been
found to add
an oscillatory component to Tassoul's asymptotic formula, but not to remove the
asymptotic structure altogether.
To treat properly the discontinuities in non-spherical stars,
the ray dynamics approach has to be extended
by taking into account the splitting of rays at the discontinuity, corresponding to the reflected
and transmitted waves. Once this is incorporated in
the formalism, quantum chaos techniques can be applied as was done for billiards in
\citet{raysplitting} and the Weyl formula can also be computed
\citep{raysplittingweyl}.

Apart from the treatment of eventual sound-speed discontinuities,
the asymptotic
analysis presented in this paper for uniformly rotating polytropic stellar models can be readily applied to realistic stellar models.
The details of the dynamics will change as they depend on the sound speed distribution of the model considered.
However,
we do not expect that the mixed character of the dynamics and thus
the irregular/regular nature of the spectrum will
change. This has to be confirmed by specific ray dynamics studies. In particular, the effects of the advection
by a differentially rotating flow should be investigated.

Lifting the two assumptions concerning the adiabaticity of the perturbations and the Coriolis force omission should
not significantly modify the results of the asymptotic analysis.
Non-adiabatic calculations are known to have a small effect on the frequency
values while the legitimate omission of the Coriolis force for high frequency motions is already relevant
for quite low frequency as shown in \citet{Rees}. 

\subsection{Implications for mode identification}

The asymptotic analysis provides qualitative and quantitative informations which can be used
to identify high frequency acoustic modes in an observed spectrum.
First the basic structure of the spectrum can be readily deduced from the ray dynamics phase space structure visualized by the PSS.
Indeed, we have seen that the $\Omega=0.59 \Omega_K$ PSS
correctly predicts that
the spectrum of axi-symmetric modes is a superposition of four frequency subsets, three regular and one irregular.
If we now look at the $\Omega=0.81 \Omega_K$ PSS, we see that the spectrum structure should be similar except that
the regular sub-spectrum associated with the 6-period island chain is no longer present
since this island chain has disappeared at this rotation rate.
In the same way, at $\Omega=0.15 \Omega_K$, we expect a simple superposition of a whispering gallery sub-spectrum
and a 2-period island mode sub-spectrum since the chaotic regions are not sufficiently developed at this rotation rate.
Such informations, while only qualitative, are crucial to guide the identification process.
Moreover this information is obtained at a relatively low computing cost since the PSS calculation is much-less demanding
than the numerical computation of modes and frequencies.

Then, the EBK quantization of the near-integrable regions provides
the values of the uniform frequency spacings (as given by Eq.~\eq{numbers}) that should be present in the observed
spectrum. When analyzing an observed spectrum, the star model is not known, thus only estimates of these uniform spacings can be obtained.
However, these estimates enable to focus the search for regularities on a smaller range of values.

Finally as we also know the frequency statistics of chaotic modes and the number of modes in each
sub-spectrum (through Weyl's formula) the asymptotic analysis enables to construct asymptotic spectra.
The chaotic mode frequencies can be obtained as a realization
of the Wigner distribution although, in this case, their frequencies could not be individually identified with observed ones.
Nevertheless, such a synthetic spectrum should be very useful to test identification methods, especially the search
for the regularities hidden in the complete spectrum (see below).

Among the additional informations which can help constraint the mode identification are the mode visibility and the mode excitation.
The excitation mechanism have been studied so far in the spherically symmetric case and need to be reconsidered for rapidly rotating stars.
The mode visibility also deserves a specific study notably the calculation of
the intensity variations induced by the oscillation in a gravity darkened atmosphere.
However, the visibility strongly depends on the cancellation effects on the disk-integrated light.
Here, we can estimate this effect by integrating the surface Lagrangian temperature perturbation of the axisymmetric modes
computed for the $\Omega=0.59 \Omega_K$ rotating polytropic star.
The disk-averaging factor is defined as:
\begin{equation}\label{eq:disk}
D(i)=
\frac{1}{\pi R_e^2 \delta T_0} \int\!\!\!\!\int_{S_v} \delta T (\theta,\phi) {\bf d S} \cdot {\bf e_i}
\end{equation}
\noi
where $i$ is the inclination angle between the line-of-sight and the rotation axis, ${\bf e_i}$ is a unit vector in the
observer's direction, $\delta T$ is the spatial part of the Lagrangian temperature perturbation
at the stellar surface and $S_v$ is the visible part of the stellar surface.
The mode amplitude is normalized by $\delta T_0$ the root mean square of the perturbation over the whole stellar surface
\begin{equation}\label{eq:t0}
\delta T_0 = \lp \frac{\int\!\!\!\!\int_{S} \delta T^2 (\theta,\phi) d S}{S} \rp^{1/2}
\end{equation}
\noi
and the projected visible surface is normalized by $\pi R_e^2$, the visible disk surface for a star seen pole-on.
With these normalizations the disk-averaging factor of an hypothetical mode uniformly distributed on the surface and seen pole-on is unity.
The method of the calculation is detailed in Appendix D and corrects the calculation described in Appendix C of \citet{Lign}
which actually only provides an approximate value of $D(i)$.

\begin{figure}
\resizebox{\hsize}{!}{\includegraphics{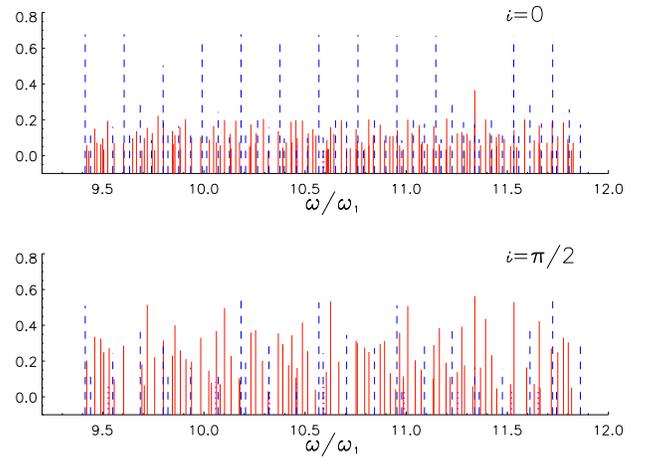}}
\caption{(Color online) Frequency spectra of axisymmetric modes where the amplitude is given by the normalized disk-averaging factor $D(i)$ for
a star seen pole-on $i =0$ and equator-on $i = \pi/2$. Only frequencies such that $D(i) \geq 2.5 \%$ are displayed and antisymmetric modes fully cancel
out equator-on. The 2-period island modes (blue/dashed lines) and the chaotic modes (red/continuous lines) are the
most visible while only a few 6-period island modes (magenta/dotted lines) and no whispering gallery mode exceed the $2.5 \%$
level.}
\label{fig.visu}
\end{figure}

Figure~\ref{fig.visu} shows the spectrum of axisymmetric modes whose disk-averaging factor exceeds $2.5$ percent. 
It appears that the disk-averaging effect does not allow to discard as many modes as for spherical stars.
Indeed, in a given frequency interval and for the same visibility threshold, we find that the number of visible modes is more than three
time higher in the $\Omega=0.59 \Omega_K$ star than in a spherical star.
Among the four classes of modes, the 2-period island modes
and the chaotic modes have similar visibilities and are significantly more visible than the 6-period island modes and 
whispering gallery modes. 
In Fig.~\ref{fig.visu}, a few 6-period island modes are visible while no whispering gallery modes exceed the chosen threshold.

The relatively high visibility of the chaotic modes with respect to the 2-period island modes was not expected
as the typical horizontal wavelength of the chaotic modes is generally significantly shorter than the one of the 2-period island modes
(see Fig.~\ref{fig.hus}). We think that this is due to the irregular nature of the node pattern of the chaotic modes
which makes the cancellation effect less effective than for regularly spaced nodes (like the whispering gallery modes).
A practical consequence of this property is that, at such a rotation rate, methods to disentangle the regular spectrum from the irregular one
should be developed. 

\subsection{Seismic constraints}

Constraints on the star can be obtained once the island and chaotic modes sub-spectra are separated.
Through the quantization formulas of the regular modes, the asymptotic analysis provides the relation between
regular frequency spacings and the physical properties of the star. For example, according to Eq.~\eq{deltas}, the seismic observable
${\delta}_n$
depends on the value of the sound velocity along the path of the 2-period periodic orbit. 
The quantity ${\delta}_{\ell}$ depends in addition 
on the second order transverse derivative of the sound velocity along the same path
and the radius of curvature of the bounding surfaces.
The 2-period periodic orbit remains along the polar axis
up to relatively high rotation rates (see Fig.~\ref{fig1} at $\Omega/\Omega_K=[0.15,0.32]$ ). This implies that all the modes trapped within the corresponding
island chain probe the center of the star which is known to be crucial for stellar evolution theory. It would be worth investigating if
high order and low degree modes of
rapidly rotating solar-type pulsators are in this case.
Other informations on the star can be deduced from the numbers of island modes or chaotic modes.
Indeed the number of mode in each class is related to the volume of the corresponding phase space
regions (see subsection 4.7) which in turns depends on the stellar model.

Further constraints on the star are expected from the identification of the chaotic modes.
Contrary to the regular modes built on invariant torus, the modes built on chaotic region are not localized in phase space and are expected 
be ergodic within their
region of propagation. This property turns out to be of particular interest for asteroseismology. The chaotic modes 
of the main chaotic region are indeed distributed all over the position space and do not avoid the stellar
core like all the non-radial modes in non-rotating stars. Thus, in rapidly rotating stars, the chaotic p-modes have the potential to probe 
the physics of the core.
While the sensitivity of the chaotic modes to this physics needs to be tested, quantum chaos studies indicate
that the spatial distribution of chaotic modes is highly sensitive to changes in the models \citep{Scha,Benenti}.

\section{Conclusion}

We have constructed the ray dynamics in uniformly rotating polytropic stars and have presented the tools and concepts
that enable to interpret it in terms of modes properties.
Accordingly, the acoustic frequency spectrum of rapidly rotating polytropic stars
is a
superposition of frequency subsets associated with
dynamically independent phase space regions. The spectra associated with near-integrable regions are regular
while those associated with chaotic regions are irregular but with specific statistical properties.
Besides this global qualitative understanding of the spectrum organization, the ray dynamics also
provides quantitative information.
The EBK quantization of the near-integrable regions enables to explicitly construct
the modes and the spectrum from the ray dynamics.
For the chaotic modes, the analysis of subsection 4.6 shows that there exists a parameter-free model of their frequency statistics.
Moreover, we have seen in subsection 4.7 that we can estimate from the ray dynamics the number of modes in each frequency subset.
These results have been confronted with the properties of acoustic modes computed in the frequency interval $[9 \omega_1, 12 \omega_1]$
showing that the present asymptotic analysis
does provide a quite accurate qualitative and quantitative understanding of the actual spectrum in this frequency range.

The analysis of section 5 argues for the importance of this asymptotic analysis
for the mode identification and for the asteroseismology of rapidly rotating stars.
Indeed, the asymptotic results and the estimation of the mode visibilities tells us
that the separation of the frequencies between chaotic and regular modes
is a necessary prerequisite in order to analyze the spectrum.
When this is done, the observed regular spacings
like $\delta_n$ and $\delta_{\ell}$ can be related 
to the internal property of the star thanks to the asymptotic analysis.

Further work on this theory could help
the analysis of the observed spectra.
First, it is important to establish more precise formulas such as
Eq.~\eq{deltas} for the regular modes
corresponding to the different stability islands.  The structure
of chaotic modes at low frequency should be studied in more details,
in frequency ranges lower than the ones used in the present work. This
will allow to test the asymptotic analysis in frequency ranges where
it is not supposed to hold in principle, but which are part of the
observed spectra. Encouragingly, the regularity of the 2-period island modes
has been already demonstrated
in a relatively low radial order range $5 \leq n \leq 10$ \citep{Lign,Rees8}, and more generally, in quantum mechanics the
semi-classical analysis has been found to apply in energy
ranges much lower than expected.
Such a study would also allow to probe if
the presence of scars (see section 4.6), which should be seen
only at low frequencies, can organize part of the
chaotic sub-spectrum.
These studies will in particular enable to produce theoretical synthetic spectra
which embody all the semi-classical information and can be used to
test methods of analysis before dealing with actual spectra.

Outside the scope of the asymptotic analysis {\it per se}, the mode identification would greatly benefit 
of accurate visibility computations, of 
modes excitation studies and obviously of more realistic models of centrifugally distorted rapidly rotating stars
\citep{Rox, MacGre, Lara}.

In conclusion, we believe that the asymptotic analysis we exposed
is a promising way to interpret the spectrum of rapidly rotating
stars. We have demonstrated that it can describe numerical spectra,
and think that with suitable refinements it should provide an important tool
for the
analysis of observed spectra such as those obtained by the instruments
COROT and KEPLER.

\begin{acknowledgements}
We thank D. Reese, M. Chapuy, S. Vidal, M. Rieutord and L. Valdettaro for their help at
various stages of this work. 
We also thank CALMIP ("CALcul en MIdi-Pyr\'en\'ees'') for the use of their supercomputer.
This work
was supported by the Programme National de Physique Stellaire of INSU/CNRS
and the SIROCO project of the Agence National de la Recherche.
\end{acknowledgements}

\appendix

\section{The WKB approximation of the stellar oscillation equations \eq{div}, \eq{vel}, \eq{adia}}

The equations are first written in a normal form, then the eikonal equation is obtained using the WKB approximation. 

\subsection{Normal form}

We first eliminate the perturbation velocity $\vec{u}$ from the equations \eq{div}, \eq{vel}, \eq{adia}
governing the evolution of the perturbations. Using equations \eq{vel}, the time derivative of Eqs.~
\eq{div} and \eq{adia}
reads:

\begin{equation} \label{rhoone}
{\partial}^2_{tt} \rho + \nab \cdot (\rho \vec{g}_0)= \Delta P,
\end{equation}

\begin{equation} \label{rhobis}
c_s^2 ({\partial}^2_{tt} + N_0^2) \rho = {\partial}^2_{tt} P + \frac{c_s^2 N_0^2}{g_0^2} 
\vec{g}_0 \cdot \nab P
\end{equation}

\noi where $N_0$ is the Brunt-V\"{a}is\"{a}l\"{a} frequency defined as 
\begin{equation} \label{Brunt}
N_0^2= \vec{g}_0 \cdot \lp \frac{\nab \rho_{0}}{\rho_{0}} - \frac{1}{\Gamma}\frac{\nab P_0}{P_0} \rp
\end{equation}

Seeking harmonic solutions in time, the variable are written $F =
\hat{F} \exp(i \omega t)$. Then, using Eq.~\eq{rhobis}, $\hat{\rho}$ is expressed in terms
of $\hat{P}$ and replaced in Eq.~\eq{rhoone} to yield:
\beqan
& \lc - \omega^4 + \omega^2 c_s^2 f \nab \cdot (\frac{\vec{g}_0}{c_s^2 f}) \rc \hat{P} + \nonumber \\
&
\lc
\omega^2 (1 + \frac{c_s^2 N_0^2}{g_0^2}) - c_s^2 f \nab \cdot (\frac{N_0^2 \vec{g}_0 }{g_0^2 f}) \rc
\vec{g}_0 \cdot \nab \hat{P}  - \nonumber \\
&  \frac{c_s^2 N_0^2}{g_0^2} \vec{g}_0 \cdot \nab (\vec{g}_0 \cdot \nab \hat{P} ) - \omega^2 c_s^2 f \Delta \hat{P} =0
\eeqan{pre}

\noi We then look for a function $\alpha$ such that the substitution $\hat{P}=  \alpha \hat{\Psi}$ eliminates
the first derivative term. The result is given by Eq.~\eq{pert} where :
\begin{equation} \label{omcgene}
\omega_c^2 = \frac{g_0^2 B^2}{4 c_s^2} + c_s^2 f \nab \cdot ( \frac{\vec{g}_0}{c_s^2 f})
- \frac{c_s^2}{2} \nab \cdot (\frac{B \vec{g}_0}{c_s^2})
+ \frac{(1-f)}{2} B g_0^2 \nab \cdot (\frac{\vec{g}_0}{g_0^2})
\end{equation}

\noi where
\begin{equation} \label{e}
B= 1 + \frac{c_s^2 N_0^2}{g_0^2} - c_s^2 \nab \cdot \lp \frac{(1-f) \vec{g}_0}{f g_0^2} \rp
\end{equation}

\noi
\begin{equation} \label{fprime}
f= 1 - \frac{N_0^2}{\omega^2}
\end{equation}

\noi The $\alpha$ function is given by:
\begin{equation} \label{alpha}
\nab \alpha = \frac{B}{2 c_s^2} \alpha \vec{g}_0 
\end{equation}

\noi The expression of $\omega_c$ can be simplified in the limit $\omega \gg N_0$ to give:
\begin{equation} \label{omcsim}
\omega_c^2 = \frac{g_0^2}{4 c_s^2} \lp 1 + \frac{c_s^2 N_0^2}{g_0^2} \rp^2
+ \frac{c_s^2}{2} \nab \cdot \lp (1 - \frac{c_s^2 N_0^2}{g_0^2}) \frac{\vec{g}_0}{c_s^2} \rp
\end{equation}
%
For polytropes $P_0 = K \varrho_0^{1+1/\mu}$, the quantities
describing the equilibrium model can be expressed in terms of the enthalpy
$h_0$ as follows:
\begin{equation} \label{eq:mod}
\begin{array}{ll}
\vec{g}_0 = \nab h_0 & c_s^2 = \frac{{\Gamma}}{\mu+1} h_0 \\
N_0^2 = \lp \frac{\mu {\Gamma}}{\mu+1} -1 \rp \frac{g_0^2}{c_s^2} & \\
\end{array}
\end{equation}

\noi Equation~\eq{omcsim} can then be simplified to give Eq.~\eq{om1dd}. We note that while
the $\omega \gg N_0$ is expected to be valid for acoustic modes in real stars, the fact that
the Brunt-V\"{a}is\"{a}l\"{a} frequency becomes infinite at the surface
of a polytropic model 
implies that this approximation is not valid close to the surface of such models.

\subsection{Eikonal equation}

We look for wave-like solutions \eq{eq:WKB} of the normal form of the perturbation equation \eq{pert} under the
assumption that $1/\Lambda$ the ratio between the wavelength of the solution and
the background typical lengthscale is very small. 
Accordingly, the solution is expanded as: 
\begin{equation} \label{exp}
\begin{array}{ll}
\Phi= \Lambda ( \Phi_0 + \frac{1}{\Lambda} \Phi_1 ..) & \;\;
A = A_0 + \frac{1}{\Lambda} A_1 ..
\end{array}
\end{equation}
\noi and the eikonal equation corresponds to the leading order of the expanded solution.

The highest $O(\Lambda^2)$ terms verify:
\begin{equation} \label{eiki} 
\frac{\omega^2 - \omega_c^2}{c_s^2} + \frac{N_0^2}{\omega^2} \Lambda^2 (\nab \Phi_0)_{\perp}^2= \Lambda^2 (\nab \Phi_0)^2
\end{equation}
\noi where the $(\nab \Phi_0)_{\perp} = \nab \Phi_0 - (\nab \Phi_0 \cdot \vec{n}_0) \vec{n}_0$,
$\vec{n}_0$ being the outward unit vector in the direction opposite to the effective gravity.
The effective eikonal equation then depends on the ordering of $\omega /\omega_0 $ with respect to $\Lambda$. 
If $\omega/\omega_0 = O(\Lambda)$, then the above equation simplifies
to:
\begin{equation}
\frac{\omega^2 - \omega_c^2}{c_s^2} = (\nab \Phi_0)^2
\end{equation}
\noi which corresponds to the dispersion relation of acoustic waves. The
$\omega_c$ term is retained because its sharp increase near the surface provokes the back reflection of the acoustic wave.

On the other hand, if $\omega /\omega_0 = O(1)$, then we obtain
\begin{equation}
-\frac{\omega_c^2}{c_s^2} + \frac{N_0^2}{\omega^2} (\nab \Phi_0)_{\perp}^2=  (\nab \Phi_0)^2
\end{equation}
\noi which corresponds to gravity waves. Note that this relation has been obtained under the
assumption that the Coriolis force is negligible. While justified for high frequency acoustic waves,
this assumption is not necessarily true for gravity waves whose frequency is limited by $N_O$.

Finally, the next order of the expansion \eq{exp} yields the amplitude $A_0$ in terms of $\Phi_0$ (see for example
\citet{Light}).

\section{Properties of the PSS}

Two specific properties of the $r_p (\theta)=r_s (\theta) - d$ PSS are considered below. First, 
we check in the non-rotating case that the distance $d$ can be chosen in order that all relevant trajectories intersect 
the $r_p (\theta)=r_s (\theta) - d$ curve. Second, we define a coordinate system of the PSS which ensures that
any surface of the PSS is conserved by the dynamics.

\subsection{Choice of the distance to the stellar surface}

In the non-rotating case, equation~\eq{pssint} enables to characterize the trajectories that do not cross the PSS for a given value of $d$,
the distance to the stellar surface.
The radius of the internal caustic of
these trajectories $r_i$ must be such that $r_i > r_p$. Using the definition of $r_i$ and assuming that
$\omega \gg \omega_c(r_i)$, implies that $\tilde{L} > r_p/c_s(r_p)$.
According to
the relation between $\tilde{L}$ and $\ell$ the degree of the corresponding mode $\tilde{L} = (\ell + 1/2)/ \omega$ (see Eq.~\eq{quant2} in
subsection 4.1),
we find that these trajectories are associated with high degree modes ($\ell > 136$) for the chosen value of
$d= 0.08 R_e$ and for $\omega = 8.4 \omega_1$, where $\omega_1$ is the lowest acoustic mode frequency of the stellar model considered.
These modes are thus irrelevant for asteroseismology since their amplitude strongly
cancels out when integrated over the visible disk.

\subsection{Area-preserving coordinates of the PSS}

The PSS being defined by $r_p (\theta)=r_s (\theta) - d$, we show here that
$\theta$, the colatitude, and $\tilde{k}_{\theta}=k_{\theta}/\omega$, $k_{\theta}$ the angular component
of $\vec{k}$
in the
natural basis associated with the coordinate system $[\zeta=r_s (\theta)-r, \theta, \phi]$,
are area-preserving coordinates of the PSS.

First we show that, for a general coordinate system ${x_i}$, the spatial coordinates ${x_i}$ and
the covariant component $\tilde{k}_i$
of the vector $\vec{\tilde{k}}$ in the natural basis are conjugate variables of the Hamiltonian $H$ given by Eq.~\eq{ham}. 
The natural basis associated to a coordinate system ${x_i}$ is defined by $(\vec{E}_1, \vec{E}_2, \vec{E}_3)$ where
$\vec{E}_i = \partial \vec{x} / \partial x^i$. The contravariant component $\tilde{k}^i$ of the velocity 
$\vec{\tilde{k}}=\vec{\dot{x}}$ thus verify $\dot{x}_i = \tilde{k}^i$ (the notation $\dot{x}_i$ denotes a derivative with
respect to the time-like coordinate $t$).
If $L(\vec{\dot{x}},\vec{x},t)$ is the Lagrangian of a system expressed in a coordinate system $x_i$, it is well
known that a Legendre transformation enables to construct a Hamiltonian
$H=\sum \dot{x}_i \partial L / \partial \dot{x}_i - L$ where $p_i=\partial L /\partial \dot{x}_i$ 
is conjugate to $x_i$.
The Lagrangian $L = \frac{\vec{\tilde{k}}^2}{2} - W(\vec{x})$ being associated with the Hamiltonian $H$ given by Eq.~\eq{ham}, 
the momentum variable $p_i$
can be simply computed:
\begin{eqnarray}
p_i=\frac{\partial L}{\partial \dot{x}_i}= \frac{1}{2} \frac{\partial \vec{\tilde{k}}\cdot \vec{\tilde{k}}}{\partial \tilde{k}^i} = \vec{\tilde{k}} \cdot \frac{\partial \vec{\tilde{k}}}{\partial
\tilde{k}^i} = \vec{\tilde{k}} \cdot \vec{E}_i =\tilde{k}_i
\end{eqnarray} 
\noi thus showing that $[x_i,\tilde{k}_i]$ are conjugate variables of $H$.

Moreover, for a given conjugate coordinate system $[x_i, p_i]$, the coordinates $[x_2,x_3,p_2,p_3]$ of the PSS defined
by $x_1=const$ are known to be area-preserving \citep{Ott}.
Thus, in our case, $[\zeta,\theta,\tilde{k}_{\zeta},\tilde{k}_{\theta}]$ is a conjugate coordinate system for the reduced Hamiltonian $H_r$
and the system $[\theta, \tilde{k}_{\theta}]$ is area-preserving for the PSS $\zeta=const$.
Note that the natural basis and its conjugate reads $\vec{E}_{\zeta}=-\vec{e}_r, \vec{E}_{\theta}=(d r_s/\!d\theta) \; \vec{e}_r + r_p \vec{e}_{\theta}$
and $\vec{E}^{\zeta}=-\vec{e}_r + [(d r_s/\!d\theta)/r_p] \vec{e}_{\theta}, \vec{E}^{\theta}= (1/r_p) \vec{e}_{\theta}$
in terms of the unit vector associated with the spherical coordinates $(\vec{e}_r, \vec{e}_{\theta})$.
Thus, with respect to the wavevector components in spherical coordinates $\tilde{k}^{sph}_r , \tilde{k}^{sph}_{\theta}$,
the $\tilde{k}_{\theta}$ component reads $\tilde{k}_{\theta} = (d r_s/\!d\theta) \; \tilde{k}^{sph}_r + r_p \tilde{k}^{sph}_{\theta}$.

\section{Calculation of phase space volumes}

Following the Monte-Carlo quadrature method \citep{Montecarlo}, $N_S$ points are randomly
chosen in a known volume $V_S$ that includes the volume $V$ to be computed.
The proportion of points
inside $V$ approximates the ratio $V/ V_S$, thus providing an estimated value of $V$.
The standard deviation error yields an estimate of the relative error, $\sqrt{(V_M/V -1)/N_S}$,
showing that the sampling volume $V_S$ has to be as close as possible to the volume $V$
and that the number of sampling points must be large.

\begin{figure}
\resizebox{\hsize}{!}{\includegraphics{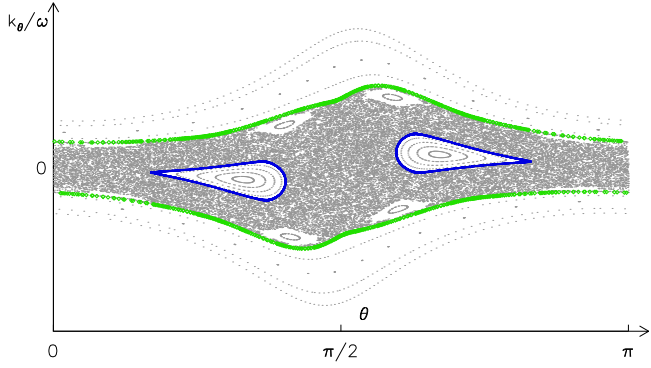}}
\caption{(Color online) Intersection of two trajectories with the PSS at $\Omega = 0.59 \Omega_K$.
The crosses (blue) correspond to a trajectory on a torus containing most of the 2-period island chain. The diamonds (green) correspond to a trajectory
on a torus bounding the central chaotic region.}
\label{fig.weyl}
\end{figure}

In our case, the main practical issue is thus to determine if a given point in phase space is inside or outside
the 4-dimensional volume to be computed.
The two volumes that we have computed are specified by two limiting frequencies $9.42 \omega_1 \leq H^{'}_r (\vec{x},\vec{k}) \leq 11.85 \omega_1$ 
and for each value of $H^{'}_r$ by the 3D volume inside a given 2D torus.
The first torus considered
separates the whispering gallery region from the chaotic region, its imprint on the $r=r_p$ PSS being shown on Fig.~\ref{fig.weyl}.
The volume inside this torus includes the large chaotic region as well as the island chains around the 2-period and 6-period orbits.
The second volume corresponds to the 2-period island chain and is delimited by a torus also shown on Fig.~\ref{fig.weyl}.

To determine whether a given point $\vec{x}_0,\vec{k}_0$ is inside or outside these volumes,
one could construct the $r=r_p$ PSS associated to the value of the
Hamiltonian $H^{'}_r (\vec{x}_0,\vec{k}_0)$, advance the dynamics from $\vec{x}_0,\vec{k}_0$ until the trajectory cross the $r=r_p$ PSS
and find out whether the crossing point is inside or outside the torus.
Here, to simplify the procedure, we used the fact that the $r=r_p$ PSS plotted against the scaled wavenumber $\vec{\tilde{k}}$
appeared to be insensitive to values of the frequency $\omega$ in the domain considered. We thus consider the point
$\vec{x}_0, \vec{k}_0 / H^{'}_r (\vec{x}_0,\vec{k}_0)$ and determine its location in the scaled phase space 
computed for a given frequency. To control this supposedly small frequency effect the computation has been performed
for the two extreme frequencies
$\omega = 9.42 \omega_1$ and $\omega=11.85 \omega_1$.
Moreover, instead of advancing the dynamics up to the $r=r_p$ PSS, we construct a local PSS (either from a $\zeta=const$
surface
or a $\tilde{k}_{\zeta}=const$ surface)
to compare the location of the $\vec{x}_0,\vec{k}_0$ point with the local imprint of the delimiting torus.
In practice, the imprint of the delimiting torus is not a continuous curve as the torus is actually obtained
from a space filling trajectory on the torus. We therefore follow such a trajectory over a sufficiently 
large number of time step to increase the number of point of the torus imprint on the different PSS.
This procedure has been tested using trajectories which are known to be either inside or outside the torus
(like for example trajectories on nested tori inside the 2-period island chain). The number of points wrongly
located by this procedure can be made small enough to have a negligible effect when compared to the statistical error on the volume determination.
Furthermore, the integration domain has been divided in three sub-domains following the pseudo-radial direction $\zeta$.
This enables to limit the ratio $V/ V_S$ as the sampling volumes can be more easily reduced in each sub-domain.

Accordingly the number of modes within the 2-period island chain is $33 \pm 1$ if we use the bounding torus computed for $\omega=9.42 \omega_1$
and $34 \pm 1$ for $\omega=11.85 \omega_1$. The effect of changing the frequency is small in this case and the number of modes
can be estimated to $34 \pm 2$.
Likewise, the number of modes outside the whispering gallery region is $263 \pm 1$ for $\omega=9.42 \omega_1$ bounding torus
and $276 \pm 2$ for $\omega=11.85 \omega_1$. The frequency effect is larger but still reasonable for the present purpose. We took the mean value of $270$ modes with an error of $\pm 8$ mode.

\section{Calculation of the disk-integration factor}

According to the definition of the disk-integration factor, Eq.~\eq{eq:disk}, we are led to calculate integrals of the following form:
\beqan
I &=& \int\!\!\!\int_{S_v} F(\theta,\phi) {\bf d S} \cdot {\bf e_i} \\
 &=& \int\!\!\!\int_{S_v} G(\theta,\phi,i) d\mu d\phi
\eeqan{aaaa}
\noi where $\mu = \cos \theta$ and  $F(\theta,\phi)=W(\theta) e^{im\phi}$
is the surface distribution of the eigenfunction. The spherical coordinate system $[r, \theta, \phi]$ is such that
the polar axis
is the rotation axis, and the meridional plane $\phi=0$ is chosen parallel to ${\bf e_i}$.
The condition ${\bf d S} \cdot {\bf e_i} =0$ on the stellar surface defines a curve
which separates the visible part of the surface $S_v$ from the invisible one
(the surface is supposed to be convex). We call this curve the visibility curve.

We used two methods to compute the integral $I$. The first one is approximate as it assumes that the visibility curve
is contained in a plane but it is easy to compute accurately. The second
method does not make this assumption but requires more computing time to complete accurate calculations.

The vector $\vec{dS}$ at the star's surface reads:
\begin{equation}
\vec{dS} =  \sin \theta r_s \left({r}_s^2 + \left( \frac{d r_s}{d \theta}\right)^2 \right)^{1/2} \vec{e^s} d\theta d\phi
\end{equation}
\noi where $\vec{e^s}$ denotes the unit vector perpendicular to the surface and $r_s(\theta)$ is the stellar surface.
Thus, the function to integrate can be written as:
\begin{equation}
F(\theta, \phi) {\bf d S} \cdot {\bf e_i} = A(\theta) \sin \theta \cos i e^{im\phi} - B(\theta) \sin \theta \sin i \cos \phi e^{im\phi}
\label{eq:F}
\end{equation}
\noi where
\beqan
A &=& r_s \frac{d}{d \theta} \lp r_s \sin\theta \rp W(\theta) \\
B &=& r_s \frac{d}{d \theta} \lp r_s \cos\theta \rp W(\theta)
\eeqan{aaaf}

\subsection{First method}

The colatitude verifying ${\bf d S} \cdot {\bf e_i} =0$ for $\phi =0$ is denoted ${\theta}_{L}(i)$. As the inclination angle $i$
can be restricted to $[0, \pi/2]$,
we have $\pi/2<{\theta}_{L}(i)<\pi$ and the angle $\alpha$ defined as $\alpha(i)={\theta}_{L}(i)-\pi/2$ verifies  $0<\alpha(i)<\pi/2$.
We assume that the visibility curve is the intersection of the stellar surface with the plane
$\sin \alpha x + \cos \alpha z = 0$, where the vector ${\bf e_{\alpha}}$ of Cartesian coordinates
$(\sin \alpha,0,\cos \alpha)$ is normal to this plane.

Then, the integral is most simply calculated in the
coordinate system in which the polar axis is aligned with the direction of the vector ${\bf e_{\alpha}}$.
This coordinate system results from a rotation of angle $\alpha$ around the $y$ axis of the
original coordinate system, the
new angular variables being denoted $\theta'$ and $\phi'$.

To express the integrand in these coordinates, we use the formula relating the spherical harmonics in both systems:
\begin{equation}
\YL(\theta,\phi) =
\sum_{m' =-\ell}^{+\ell} d_{m m'}^{\ell}(\alpha) \YLM(\theta ',\phi ')
\end{equation}
\noi where $d_{m m'}^{\ell}(\alpha)$ do not generally have a simple form (Edmonds 1960).
Then, using the spherical harmonic expansion of $G$, we obtain:
\beqan
G &=& \sum_{\ell=0}^{+\infty}\sum_{m=-\ell}^{+\ell} G^\ell_m(i) \YL(\theta,\phi) \\
 &=& \sum_{\ell=0}^{+\infty}\sum_{m=-\ell}^{+\ell} \sum_{m'=-\ell}^{+\ell} G^\ell_m(i) d_{m m'}^{\ell}(\alpha) \YLM(\theta',\phi')
\eeqan{aaab}
Then, integrating over the longitude $\phi'$, from $0$ to $2 \pi$, the terms
involving $\YLM(\theta',\phi')$ vanish if $m' \neq 0$. It follows that
\begin{equation}
I = 2 \pi \sum_{\ell=0}^{+\infty}\sum_{m=-\ell}^{+\ell} J_{\ell} G^\ell_m(i) \YTL(\alpha)
\end{equation}
\noi where we used the following relations,
\begin{equation}
d_{m 0}^{\ell}(\alpha) = \sqrt{\frac{4 \pi}{2 \ell + 1}} \YTL(\alpha)
\end{equation}
\begin{equation}
d\mu d\phi =  d\mu' d\phi' \;\; \mbox{where $\mu' = \cos \theta'$}
\end{equation}
\noi and defined $J_{\ell}$ as,
\beqan
J_{\ell} &=& \sqrt{\frac{4 \pi}{2\ell+1}} \int_0^{1} \hat{Y}^0_\ell(\mu') d\mu' \\
&=& \left\{ \begin{array}{lll}
         0 & \mbox{if $\ell$ is even and $\ell \neq 0$},\\
         1 & \mbox{if $\ell=0$},\\
         (-1)^{\frac{\ell-1}{2}} \frac{1.3...(\ell-2)}{2.4...(\ell+1)} & \mbox{if $\ell$ is odd and $\ell \neq 1$}, \\
         \frac{1}{2} & \mbox{if $\ell=1$}.\end{array} \right.
\eeqan{aaac}

To determine the coefficients $G^\ell_m(i)$, we use the expression of $G$ derived from Eq.~\eq{eq:F}:
\beqan
G &=& A(\theta) \cos i e^{im\phi}  - B(\theta) \sin i \cos \phi e^{im\phi}
\eeqan{aaae}
\noi It follows that
\begin{equation}
G^\ell_k = 0 \;\;  \mbox{if $k \neq m-1,m,m+1$}
\end{equation}
\noi so that the integral now reads:
\beqan
I/2\pi &=& I_{m-1}+  I_m + I_{m+1} \;\; \mbox{where} \\
I_m &=& \cos i \hat{A}_m(\alpha) \\
I_{m-1} &=& - \frac{\sin i}{2} \hat{B}_{m-1}(\alpha) \\
I_{m+1} &=& - \frac{\sin i}{2} \hat{B}_{m+1}(\alpha)
\eeqan{aaag}
where $\hat{A}_m$ denotes:
\beqan
\hat{A}_m(\alpha) &=& \sum_{\ell=|m|}^{+\infty} J_\ell A^\ell_m \YTL(\alpha) \\
A^\ell_m &=& 2\pi \int_0^{\pi} A(\theta) \YTL(\theta) \sin \theta d \theta
\eeqan{aaah}
\noi the $\hat{B}_m$ terms being defined accordingly.

Note that for modes which are equatorially anti-symmetric and axisymmetric ($m=0$),
$\hat{A}_0(\alpha)=J_0 A^0_0 \hat{Y}^0_0(\alpha)$ and $\hat{B}_1(\alpha) = \hat{B}_{-1}(\alpha) = 0$, thus
the integral $I$ reduces to:
\beqan
I = 4 \pi \sqrt{\pi} A^0_0 \cos(i).
\eeqan{aaai}

\subsection{Second method}

The visibility curve is no longer assumed to be planar. The integration over the visible surface is first performed in the azimuthal
direction and then in latitude. If $0 \le \theta \le \pi/2 - \alpha$ , the integration is made between $0$ and $2 \pi$, while in the interval
$\pi/2 - \alpha \le \theta \le \pi/2 + \alpha$ one has to integrate between the two limiting azimuths $-{\phi}_L(\theta, i)$
and ${\phi}_L(\theta, i)$ 
verifying ${\bf d S} \cdot {\bf e_i} =0$.
The integration domain is thus divided in two sub-domains, such that
\begin{equation}
S_v = [0, \frac{\pi}{2}-\alpha ] \times [0,2 \pi ] \cup [ \frac{\pi}{2}-\alpha,\frac{\pi}{2} +\alpha ] \times [ -{\phi}_L,{\phi}_L]
\end{equation}

According to Eq.~\eq{eq:F}, the integration over $\phi$ can be made
analytically as it involves quadratures of $e^{im\phi}$ and $\cos \phi e^{im\phi}$ over
$[0,2\pi]$ and $[-{\phi}_L,{\phi}_L]$. Then, depending on the value of $m$, the integral $I$ reads:
\begin{equation}
I = \left\{ \begin{array}{l}
         2\pi\int_{0}^{\frac{\pi}{2}-\alpha} a d\theta + 2 \int_{\frac{\pi}{2}-\alpha}^{\frac{\pi}{2}+\alpha} a {\phi}_L -b \sin({\phi}_L) d\theta \;\; \mbox{if $m = 0$}\\
        -\pi\int_{0}^{\frac{\pi}{2}-\alpha} b d\theta + \int_{\frac{\pi}{2}-\alpha}^{\frac{\pi}{2}+\alpha} 2a \sin({\phi}_L) -b\left({\phi}_L + \frac{1}{2} \sin(2{\phi}_L)\right) d\theta \\
\;\;\;\;\;\;\;\;\; \mbox{if $m = \pm 1$} \\
        \int_{\frac{\pi}{2}-\alpha}^{\frac{\pi}{2}+\alpha} \frac{2a}{m} \sin (m {\phi}_L) -b\left(\frac{\sin [(m+1) {\phi}_L]}{m+1} + \frac{\sin [(m-1) {\phi}_L]}{m-1}\right) d\theta \\
\;\;\;\;\;\;\;\;\; \mbox{if $m \ne 0,\pm 1$} .\end{array} \right.
\end{equation}

\noi where $a(\theta,i)= A(\theta) \sin \theta \cos i$ and $b(\theta,i)=B(\theta) \sin \theta \sin i$.

\subsection{Tests}
The methods have been tested in the case of an uniformly distributed function on the surface
of an ellipso\"id where they should both give the same result. 
Then, the error introduced by the approximation of method $1$ is estimated in the case of a Roche model surface.

\subsubsection{Ellipso\"id}
The surface being a quadric, the visibility curve is planar. Method $1$ is therefore exact and should give the same result
as 
method $2$. In addition, the visible surface (obtained by taking $F=1$)
can be obtained analytically.
Indeed, the dimensionless equation of the ellipso\"id is:
\begin{equation}
x^2 + y^2 + \frac{z^2}{\tilde{R}_p} = 1
\end{equation}
and in spherical coordinates:
\begin{equation}
r=\frac{1}{\sqrt{1+\frac{1-\tilde{R}_p^2}{\tilde{R}_p^2} \cos^2 \theta}}
\label{eq:ellips}
\end{equation}
\noi where the distance have been normalized by the equatorial radius $R_e$ and $\tilde{R}_p = R_p / R_e$.
As ${\bf d S}$ is parallel to $(2x,2y,2z/\tilde{R}_p^2)$ and ${\bf e_i} = (\sin i,0,\cos i)$, surface points verifying ${\bf d S} \cdot {\bf e_i} =0$ belong to
the plane:
\beq
x \sin i + \frac{z \cos i}{\tilde{R}_p^2} = 0
\label{eq:plan}
\eeq
By definition of the angle $\alpha$, we have
\beq
\tan \alpha = \tilde{R}_p^2 \tan i
\label{eq:alpha}
\eeq
In addition, using Eq.~\eq{eq:plan} and the relation between
the Cartesian and spherical coordinates, the equation of the intersection
between the plane $x \sin i + (z \cos i)/\tilde{R}_p^2 = 0$ and the ellispso\"id
is given by Eq.~\eq{eq:ellips} and
\beq
\cos \phi_L= - \frac{\cot \theta \cot i}{\tilde{R}_p^2}= - \cot \theta \cot \alpha
\label{eq:phil}
\eeq
Thus,
\beq
\phi_L (\theta, i) = \arccos (- \cot \theta \cot \alpha)
\eeq
is defined if
$\pi/2 - \alpha \le \theta \le \pi/2 + \alpha$ and verifies $0 \le {\phi}_L \le \pi$.

The visible surface can be calculated analytically as the visible curve is an ellipse.
This can be seen
using the Cartesian coordinates obtained by the rotation of angle $\alpha$ about the $Oy$
axis:
\beqan
x'& = & \cos \alpha x  - \sin \alpha z \\
y' & = & y \\
z' & = & \sin \alpha x + \cos \alpha z
\eeqan{rotxyz}

The curve is then contained in the plane $z'=0$ and verifies:
\beq
(\cos^2 \alpha + \frac{\sin^2 \alpha}{\tilde{R}_p^2}) x'^2 + y'^2 = 1
\eeq
The surface of this ellipse can be calculated as well as its projection
in the direction ${\bf e_i}$, denoted $S^p_v$ :
\beqan
S^p_v / R_e^2 & = & \pi \cos (\alpha - i) \sqrt{\frac{1+\tan^2 i \tilde{R}_p^4}{1+\tan^2 i \tilde{R}_p^2}}\\
& = & \pi \cos i \sqrt{1+\tan^2 i \tilde{R}_p^2}
\eeqan{svellips}

Both method were successfully tested against this analytical expression. 
Method $1$ is simpler as it does not require a numerical integration.
It is also very accurate although, if one wants to reach machine precision, it is necessary to use the analytical value of $\theta_L$.

\subsubsection{Method 1 versus Method 2}
\begin{figure}[h]
\resizebox{\hsize}{!}{\includegraphics{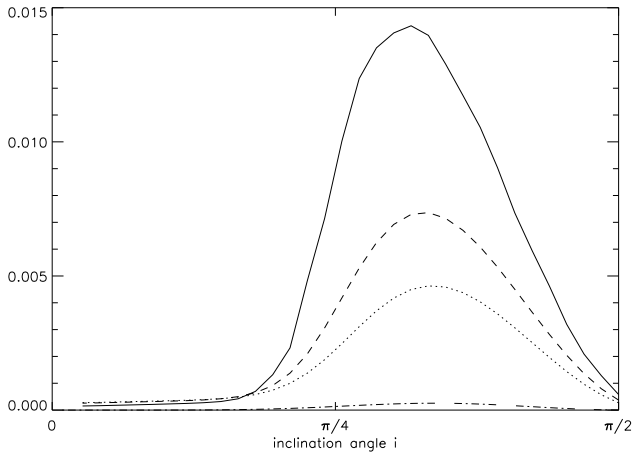}}
\caption{Relative error of the projected visible surface computed with method $1$
for Roche models of different flatness : 0.1526 (dot-dashed), 0.2594 (dotted) 0.2804 (dashed), 0.3092 (continuous line).}
\label{er_int}
\end{figure}

Method $1$ is approximate as it assumes that the curve on the surface verifying ${\bf d S} \cdot {\bf e_i} =0$
is planar and contained in the plane $\sin \alpha x + \cos \alpha z = 0$. Nevertheless, the associated
error is not expected to be large as it concerns a small part of the whole visible surface.
To test this, we simply compute the integral for $F=1$ with both methods for a surface given by a Roche model
of rotating star.
Method $2$ computes the projected visible surface while the quantity calculated by method $1$ is smaller
as the integration includes a region where ${\bf d S} \cdot {\bf e_i} < 0$ and excludes
a visible region of equivalent surface which is symmetrical with respect to the star center.
The difference between the two quantities can be also calculated by directly computing the integral:
\beqan
2 \int_{\frac{\pi}{2}}^{\frac{\pi}{2}+\alpha} a {\phi}_L -b \sin({\phi}_L) d\theta
\eeqan{dile}
using either the correct value of ${\phi}_L$ or the one corresponding
to the assumption of method $1$, that is
\beqan
\cos {\phi'}_L= - \cot \theta \cot \alpha
\eeqan{dole}
In Fig.~\ref{er_int}, the relative error on the projected visible surface due to method $1$ is plotted for Roche model surfaces of different
flatness. 

It appears that, except for the near critical values of the flatness, the visible surface that is not considered
by method $1$ is a very small fraction of the total visible surface. Using method $1$ is therefore a good approximation
in these cases. For near critical flatness, the difference remains small although it can be useful to test
results of method $1$ with method $2$.

\end{document}